\documentclass[twocolumn,showpacs,preprintnumbers,amsmath,amssymb,floatfix,nofootinbib]{revtex4-1}
\usepackage{graphicx,color}
\usepackage{dcolumn}
\usepackage{bm}
\usepackage{amsthm}
\usepackage{amsmath}
\usepackage{amsfonts}

\newcommand{\be}{\begin{equation}}
\newcommand{\ee}{\end{equation}}
\newcommand{\ba}{\begin{eqnarray}}
\newcommand{\ea}{\end{eqnarray}}
\newcommand{\hhh}{\,,\hspace{0.2cm}}
\newcommand{\non}{\nonumber}
\newcommand{\n}[1]{\label{#1}}
\newcommand{\eq}[1]{(\ref{#1})}

\newcommand{\tu}{\widetilde{U}}
\newcommand{\tw}{\widetilde{W}}
\newcommand{\tv}{\widetilde{V}}
\newcommand{\tnu}{\widetilde{\nu}}
\newcommand{\hu}{\widehat{U}}
\newcommand{\hw}{\widehat{W}}
\newcommand{\hx}{\widehat{X}}
\newcommand{\hv}{\widehat{V}}
\newcommand{\hnu}{\widehat{\nu}}
\newcommand{\cu}{{\cal U}}
\newcommand{\cw}{{\cal W}}
\newcommand{\cv}{{\cal V}}

\newcommand{\ind}[1]{\mbox{\tiny #1}}

\begin{document}
 
\title{Distorted 5-dimensional vacuum black hole} 

\author{Shohreh Abdolrahimi}
\email{abdolrah@ualberta.ca}
\author{Andrey A. Shoom}
\email{ashoom@ualberta.ca}
\author{Don N. Page}
\email{profdonpage@gmail.com}
\affiliation{Theoretical Physics Institute, University of Alberta, 
Edmonton, AB, Canada,  T6G 2G7}
\date{\today}

\begin{abstract}

In this paper we study how the distortion generated by a static and neutral distribution of external matter affects a 5-dimensional Schwarzschild-Tangherlini black hole. A solution representing a particular class of such distorted black holes admits an $\mathbb{R}^{1}\times U(1)\times U(1)$ isometry group. We show that there exists a certain duality transformation between the black hole horizon and a {\em stretched singularity} surfaces. The space-time near the distorted black hole singularity has the same topology and Kasner exponents as those of a 5-dimensional Schwarzschild-Tangherlini black hole. We calculate the maximal proper time of free fall of a test particle from the distorted black hole horizon to its singularity and find that, depending on the distortion, it can be less, equal to, or greater than that of a Schwarzschild-Tangherlini black hole of the same horizon area. This implies that due to the distortion, the singularity of a Schwarzschild-Tangherlini black hole can come close to its horizon. A relation between the Kretschmann scalar calculated on the horizon of a 5-dimensional static, asymmetric, distorted black hole and the trace of the square of the Ricci tensor of the horizon surface is derived.
 
\end{abstract}

\pacs{04.50.Gh, 04.70.Bw, 04.20.Jb \hfill  
Alberta-Thy-12-10}

\maketitle

\section{INTRODUCTION}

String theories, the AdS/CFT correspondence \cite{Maldacena,AdS}, the ADD model \cite{ADD1,ADD2}, and brane-world RS models \cite{RS} suggest that higher-dimensional solutions of general relativity may have physical applications. Whether our world has extra dimensions or not should be eventually verified by experiments. One of such experiments is microscopic black hole production, which may be conducted at the LHC. Such a black hole may be created at energies of the order of $\sim 10$ TeV, if our world has extra spatial dimensions of large size ($<1$ mm) or large warping, which become accessible on such energetic scales (see, e.g., \cite{DL,GT,Kanti1,YM,GM,Kanti2,Ruth,Gingrich}).  

Possible physical applications of higher dimensions have increased interest in higher-dimensional solutions of general relativity. However, the Einstein equations of general relativity, especially higher-dimensional ones, are very complex. To solve them we have to use numerical computations, except for some idealized, highly-symmetrical cases, when construction of analytical solutions becomes possible. For example, one such construction, corresponding to a 4-dimensional, static and axisymmetric vacuum space-time, is due to Weyl \cite{Weyl}. The Weyl solution implies a static and axisymmetric distribution of matter. One of the Einstein equations for the space-time metric represented in the Weyl form reduces to a linear Laplace equation. Therefore, the superposition principle can be applied for the construction of one of the metric functions. Another metric function can be derived by a line integral in terms of the first one. As a result, one can relatively easily construct many interesting solutions, e.g., the Israel-Khan solution representing a set of collinear Schwarzschild black holes \cite{IsKh}, a black hole with a toroidal horizon \cite{Pet}, and a compactified black hole \cite{BP,FF,May}.

In higher-dimensional space-times we have a very rich variety of black objects classified according to their horizon topology, for example black holes, black strings, and black rings (for a review see, e.g., \cite{ER}). However, an exact analytical solution representing a black hole in a space-time with one large, compact extra dimension is not known. The solution representing a black hole in a 5-dimensional space-time with one large, compact extra dimension is not algebraically special \cite{Smet}. As a result, finding such analytical solution can be a formidable problem. Analytical approximations to the black hole are given in \cite{HO,H,Dan1,Dan2,Karasik}. Finding a solution representing a black hole localized on a brane is not a simple problem either. A numerical analysis suggests that in a 5-dimensional, one-brane RS model, only a black hole whose horizon radius is smaller than the bulk curvature can be localized on the brane \cite{Kudoh}. Results of a subsequent numerical analysis further suggest that such a black hole may be unstable \cite{Taka}.

Both the sought black hole solutions are axisymmetric, in the sense that they admit an $SO(3)$ isometry group. Orbits of the group are 2-dimensional spheres of nonzero curvature. As was noticed in \cite{May}, this nonzero curvature is an essential problem for a construction of such higher-dimensional axisymmetric solutions. However, one can construct algebraically special axisymmetric solutions in $d$-dimensional space-times \cite{MR}. As it was  concluded in \cite{Emp}, a $d$-dimensional, axisymmetric space-time which admits the $SO(d-2)$ isometry group cannot be considered as an appropriate higher-dimensional generalization of the 4-dimensional Weyl form. Instead, it was proposed in \cite{Emp} to consider a $d$-dimensional space-time which admits the $\mathbb{R}^{1}\times U(1)^{d-3}$ isometry group. Such a generalized Weyl form allows for the construction of many interesting black objects (see, e.g., \cite{ER,Emp}). However, as it was illustrated in \cite{ER}, only 4- and 5-dimensional black hole solutions can be presented in the Weyl form. Let us mention that a generalization of the Weyl form to the Einstein-Gauss-Bonnet theory in a 5-dimensional space-time was proposed in \cite{KKR}. Numerical evidence that a Schwarzschild black hole, a static black ring, and a uniform black string can also be considered within the generalization of the Weyl form was given in \cite{BD,Wheel}.

Having the generalized Weyl form, one may try to construct higher-dimensional analogues of 4-dimensional Weyl solutions. For example, a construction of multi-black-hole configurations within the generalized 5-dimensional Weyl form was discussed in the paper \cite{ER}. The first configuration discussed there represents a two-black-hole solution which is not asymptotically flat. The second configuration is a three-black-hole solution which is asymptotically flat but suffers from irremovable conical singularities. In addition, the central black hole is collinear with the other two along {\em different} symmetry axes. The third configuration represents an infinite number of black holes. However, it does not correspond to a 5-dimensional {\em compactified} black hole (``caged'' in the compact dimension). In fact, such a black hole corresponds to a space-time with an infinite number of collinear black holes which admits an $\mathbb{R}^{1}\times SO(3)$ isometry group, instead of the $\mathbb{R}^{1}\times U(1)\times U(1)$ isometry of the 5-dimensional Weyl form. Asymptotically flat space-times which admit an $\mathbb{R}^{1}\times U(1)\times U(1)$ isometry and correspond to 5-dimensional ``collinear'' black holes  were constructed in \cite{TT}. The corresponding background space-times have conical singularities and are not flat by the construction. Such space-times have more than one fixed point of the $U(1)\times U(1)$ isometry group, whereas a 5-dimensional Minkowski space-time has only one such point. 

Black holes interact with external matter and fields. For example, an accretion disk around a black hole tidally distorts its horizon. An accretion scenario of a black hole which may be produced at the LHC is given in \cite{GM,Cas}. As it is for any physical objects, properties of black holes are mostly revealed by their interactions. To analyze a black hole interaction is a formidable problem which requires involved numerical computations. However, a study of idealized, highly-symmetrical analytical solutions may provide us with an exact description of black-hole nonlinear interactions with external matter and fields. Among such solutions a black hole distorted by an external, static and axisymmetric distribution of matter deserves particular attention. Such a black hole was analyzed in the papers \cite{Geroch,Chandrabook,FN,Dor,MySz,Werner,FS}.

External matter affects the internal structure of black holes as well. For example, external, asymmetric dynamical distortion of a black hole results in chaotic and oscillatory space-time singularity of the BKL-type, which corresponds to shifts between different Kasner regimes (see, e.g., \cite{BKL,Lan,Misner}). The interior of a 4-dimensional, distorted, static and axisymmetric, vacuum black hole was studied in \cite{FS}. It was shown that in the vicinity of the black hole singularity the space-time has the same Kasner exponents as that of a Schwarzschild black hole. However, the static and axisymmetric distortion does change the geometry of the black hole {\em stretched singularity} (region near a black hole singularity where the space-time curvature is of the Planckian order, $\sim 10^{66}$ $\text{cm}^{-2}$) and horizon surfaces. The change is such that a certain duality transformation between the geometry of the horizon and the stretched singularity surfaces holds. According to that relation, the geometry of the horizon surface uniquely defines the geometry of the stretched singularity surface. In addition, it was shown that such a distortion noticeably affects the proper time of free fall from the black hole horizon to its singularity. It is interesting to study whether a higher-dimensional distorted black hole has similar properties. 

Another motivation to analyze the interior of a higher-dimensional distorted black hole is related to analysis of a topological phase transition between a nonuniform black string, whose horizon wraps the space-time compact dimension, and a compactified black hole (see, e.g., \cite{Kol,HNO}). In such a transition the black string and black hole topological phases meet at the merger point \cite{Kol,Wise,Kol2,KuWise,KK}. As a result, their near horizon geometry gets distorted. The interior of a nonuniform 6-dimensional black string was studied in \cite{KK1}, where numerical evidence of a space-time singularity approaching the black string horizon at the merger point was presented. What happens to the corresponding compactified black hole approaching the merger point and which way it gets distorted remains an interesting open question.

The main goal of our paper is to study a 5-dimensional, distorted, static, vacuum black hole as a distorted Schwarzschild-Tangherlini black hole, which can be presented in the generalized Weyl form, and to compare its properties with those of a 4-dimensional, distorted, static and axisymmetric, vacuum black hole. A 5-dimensional Schwarzschild-Tangherlini black hole is a good approximation to a 5-dimensional compactified black hole if the size of the compact dimension is much larger than the size of the black hole. Thus, the distorted Schwarzschild-Tangherlini black hole may be also considered as a good approximation for such distorted compactified black hole.  

Our paper is organized as follows: In Section II, we construct the 5-dimensional Weyl solution which includes gravitational distortion fields due to remote matter. In Section III, we present the metric of a 5-dimensional, static, vacuum black hole distorted by external gravitational fields and derive the corresponding Einstein equations. A solution to the Einstein equations is derived in Section IV. In Section V, we study the symmetry properties of the distortion fields and present their boundary values on the black hole horizon, singularity, and on its symmetry axes. The space-time near the black hole horizon and singularity is analyzed in Sections. VI and VII, respectively. In Section VIII, we discuss how the black hole distortion affects the maximal proper time of free fall of a test particle moving from the black hole horizon to its singularity. We summarize and discuss our results in Section IX. Details of our calculations are presented in the appendices. 

In this paper we use the following convention of units: $G_{(5)}=c=1$, the space-time signature is $+3$, and the sign conventions are that adopted in \cite{MTW}.

\section{5-dimensional Weyl solution}

In this Section we present a 5-dimensional generalization of the Weyl solution in the form suitable for analysis of a distorted 5-dimensional vacuum black hole. To begin with, let us briefly discuss the main properties of the 4-dimensional Weyl solution presented in the following Weyl form:
\ba\n{1.0}
&&\hspace{-0.5cm}ds^2=-e^{2U}dt^2+e^{2(V-U)}(dz^2+d\rho^2)+\rho^2e^{-2U}d\phi^2\,,
\ea
where $t,z\in(-\infty,\infty)$, $\rho\in(0,\infty)$, and $\phi\in[0,2\pi)$. The metric functions $U$ and $V$ depend on the cylindrical coordinates $\rho$ and $z$. The Weyl solution represents a general static and axisymmetric metric which solves the corresponding vacuum Einstein equations. One of these equations reduces to the following linear equation for the metric function $U$:  
\be\n{1.0a}
U_{,\rho\rho}+\frac{1}{\rho}U_{,\rho}+U_{,zz}=0\,,
\ee
which is defined on the plane $(\rho,z)$. Here and in what follows, $(...)_{,a}$ stands for the partial derivative of the expression $(...)$ with respect to the coordinate $x^a$. Equation \eq{1.0a} can be viewed as a 3-dimensional Laplace equation defined in an auxiliary 3-dimensional Euclidean space. The remaining Einstein equations define the metric function $V$ as follows:
\ba
V_{,\rho}&=&\rho\left(U_{,\rho}^2-U_{,z}^2\right)\,,\n{1.0b}\\
V_{,z}&=&2\rho\, U_{,\rho}U_{,z}\,.\n{1.0c}
\ea
Equation \eq{1.0a} is the integrability condition for Eqs. \eq{1.0b} and \eq{1.0c}. If we solve Eq. \eq{1.0a} for the metric function $U$, then the second metric function $V$ can be derived by the following line integral:
\be\n{1.0d}
V(\rho,z)=\int_{(\rho_0,z_0)}^{(\rho,z)}\left[V_{,\rho'}(\rho',z')d\rho'+V_{,z'}(\rho',z')dz'\right]\,,
\ee
where the integral is taken along any path connecting the points $(\rho_0,z_0)$ and $(\rho,z)$. The constant of integration is defined by a point $(\rho_0,z_0)$. 

The 4-dimensional Weyl solution admits an $\mathbb{R}^1_t\times SO(2)\cong\mathbb{R}^1_t\times U_{\phi}(1)$ isometry group. In other words, the Weyl solution is characterized by the two orthogonal, commuting Killing vectors $\xi^\alpha_{(t)}=\delta^\alpha_{\,\,\, t}$ and $\xi^\alpha_{(\phi)}=\delta^\alpha_{\,\,\, \phi}$, which are generators of time translations and 2-dimensional rotations about the symmetry axis $z$, respectively. Note that the metric function $U$ together with the constant of integration in \eq{1.0d} uniquely define the space-time geometry. 

The $d$-dimensional generalization of the Weyl solution which admits $d-2$ commuting, non-null, orthogonal Killing vector fields was presented in the papers \cite{Emp} and \cite{ER}.  Here we discuss the 5-dimensional generalized Weyl solution which is characterized by three commuting, non-null, orthogonal Killing vector fields, one of which ($\xi^\alpha_{(t)}=\delta^\alpha_{\,\,\, t}$) is timelike, and other two ($\xi^\alpha_{(\chi)}=\delta^\alpha_{\,\,\, \chi}$ and $\xi^\alpha_{(\phi)}=\delta^\alpha_{\,\,\, \phi}$) are spacelike. The Killing vectors are generators of the isometry group $\mathbb{R}^1_t\times U_\chi(1)\times U_\phi(1)$. Thus, the 5-dimensional Weyl solution can be presented as follows:
\ba\n{1.1}
ds^2&=&-e^{2U_1}dt^2+e^{2\nu}(dz^2+d\rho^2)+e^{2U_2}d\chi^2+e^{2U_3}d\phi^2\,,\non\\
\ea
where $t,z\in(-\infty,\infty)$, $\rho\in(0,\infty)$, and $\chi,\phi\in[0,2\pi)$. The metric functions $U_i$, $i=1,2,3$, and $\nu$ depend on the coordinates $\rho$ and $z$. Each of the functions $U_i$ solves the 3-dimensional Laplace equation \eq{1.0a} with the following constraint:
\be\n{1.3}
U_1+U_2+U_3=\ln \rho\,.
\ee

If the functions $U_i$ are known, the function $\nu$ can be derived by the line integral \eq{1.0d} using the following expressions:
\ba
\nu_{,\rho}&=&-\rho(U_{1,\rho}U_{2,\rho}+U_{1,\rho}U_{3,\rho}+U_{2,\rho}U_{3,\rho}\non\\
&-&U_{1,z}U_{2,z}-U_{1,z}U_{3,z}-U_{2,z}U_{3,z})\,,\n{1.4a}\\
\nu_{,z}&=&-\rho(U_{1,\rho}U_{2,z}+U_{1,\rho}U_{3,z}+U_{2,\rho}U_{3,z}\non\\
&+&U_{1,z}U_{2,\rho}+U_{1,z}U_{3,\rho}+U_{2,z}U_{3,\rho})\,.\n{1.4b}
\ea
The structure of the 5-dimensional Weyl solution can be understood as follows: Given three solutions $U_{i}$ of the Laplace equation \eq{1.0a} which satisfy the constraint \eq{1.3}, then norms of the Killing vectors are defined, and with the choice of the integration constant in the line integral for the function $\nu$ the space-time geometry is constructed. Because Eq. \eq{1.0a} for the metric functions $U_{i}$ is linear, the superposition principle can be applied for their construction.

Here we shall consider a 5-dimensional Weyl solution representing a background Weyl solution defined by $\tu_i$ and $\tnu$, which is distorted by the external, static, axisymmetric fields defined by $\hu_i$ and $\hnu$. The metric functions of the corresponding space-time are 
\be\n{1.5}
U_i=\tu_i+\hu_{i}\hhh \nu=\tnu+\hnu\,,
\ee
where according to the constraint \eq{1.3}, we have
\be\n{1.6}
\tu_1+\tu_2+\tu_3=\ln \rho\hhh \hu_1+\hu_2+\hu_3=0\,.
\ee
In what follows, we shall consider static distortion due to the external gravitational fields of remote masses whose configuration obeys the spatial symmetry of $U_\chi(1)\times U_\phi(1)$. Accordingly, we define
\ba
&&\hspace{-0.7cm}\tu_1:=\tu+\tw+\ln\rho\hhh\tu_2:=-\tw\hhh\tu_3:=-\tu\,,\n{1.7a}\\
&&\hspace{1.5cm}\tnu:=\tv+\tu+\tw\,,\n{1.7b}\\
&&\hspace{-0.3cm}\hu_1:=\hu+\hw\hhh\hu_2:=-\hw\hhh\hu_3:=-\hu\,,\n{1.7c}\\
&&\hspace{1.5cm}\hnu:=\hv+\hu+\hw\,.\n{1.7d}
\ea
Here the distortion fields $\hu$ and $\hv$ define the external gravitational fields, and $\hv$ defines the interaction between the fields themselves and the background space-time. Then, the metric \eq{1.1} takes the following generalized Weyl form\footnote[1]{The factor $\rho^{2}$ in $g_{tt}$ is a result of the definition of the metric functions. It can be removed by specifying their explicit form. For example, the 5-dimensional flat space-time
\be
ds^2=-dt^2+dx^2+dy^2+x^2d\phi^2+y^2d\chi^2\,\non
\ee
can be derived from the metric \eq{1.8} by taking $\hu=\hw=\hv=0$ and using the following metric functions:
\be
\tu=-\ln|x|\hhh \tw=-\ln|y|\hhh \tv=\ln\left|\frac{xy}{x^2+y^2}\right|\,,\non\\
\ee    
where $x^2=\sqrt{\rho^2+z^2}-z$ and $y^2=\sqrt{\rho^2+z^2}+z$.}:
\ba\n{1.8}
ds^2&=&e^{2(\tu+\tw+\hu+\hw)}[-\rho^2dt^2+e^{2(\tv+\hv)}(dz^2+d\rho^2)]\non\\
&+&e^{-2(\tw+\hw)}d\chi^2+e^{-2(\tu+\hu)}d\phi^2\,.
\ea
The background fields $\tu$ and $\tw$ satisfy the 3-dimensional Laplace equation \eq{1.0a}, and the function $\tv$ can be derived by the line integral \eq{1.0d} using the expressions
\ba
\tv_{,\rho}&=&\rho\,(\tu_{,\rho}^2+\tw_{,\rho}^2+\tu_{,\rho}\tw_{,\rho}-\tu_{,z}^2-\tw_{,z}^2-\tu_{,z}\tw_{,z})\,,\non\\
\n{1.9a}\\
\tv_{,z}&=&\rho\,(2\tu_{,\rho}\tu_{,z}+2\tw_{,\rho}\tw_{,z}+\tu_{,\rho}\tw_{,z}+\tu_{,z}\tw_{,\rho})\,.\n{1.9b}
\ea
The distortion fields $\hu$ and $\hw$ satisfy the 3-dimensional Laplace equation \eq{1.0a}, and the function $\hv$ can be derived by the line integral \eq{1.0d} using the expressions
\ba
\hv_{,\rho}&=&\rho\,(\hu_{,\rho}^2+\hw_{,\rho}^2+\hu_{,\rho}\hw_{,\rho}-\hu_{,z}^2-\hw_{,z}^2-\hu_{,z}\hw_{,z}\non\\
&+&\tu_{,\rho}\hw_{,\rho}+\tw_{,\rho}\hu_{,\rho}-\tu_{,z}\hw_{,z}-\tw_{,z}\hu_{,z}\non\\
&+&2[\tu_{,\rho}\hu_{,\rho}+\tw_{,\rho}\hw_{,\rho}-\tu_{,z}\hu_{,z}-\tw_{,z}\hw_{,z}])\,,
\n{1.10a}\\
\hv_{,z}&=&\rho\,(2\hu_{,\rho}\hu_{,z}+2\hw_{,\rho}\hw_{,z}+\hu_{,\rho}\hw_{,z}+\hu_{,z}\hw_{,\rho})\non\\
&+&\tu_{,\rho}\hw_{,z}+\tu_{,z}\hw_{,\rho}+\tw_{,\rho}\hu_{,z}+\tw_{,z}\hu_{,\rho}\non\\
&+&2[\tu_{,\rho}\hu_{,z}+\tu_{,z}\hu_{,\rho}+\tw_{,\rho}\hw_{,z}+\tw_{,z}\hw_{,\rho}])
\,.\n{1.10b}
\ea

In the following Sections we construct the metric representing a 5-dimensional distorted Schwarzschild-Tangherlini black hole and study its properties.
 
\section{Distorted 5-dimensional vacuum black hole}

\subsection{5-dimensional Schwarzschild-Tangherlini black hole}

A 5-dimensional Schwarzschild-Tangherlini black hole \cite{Tan} is given by the following metric: 
\be\n{2.1}
ds^2=-\left(1-\frac{r_o^2}{r^2}\right)dt^2+\left(1-\frac{r_o^2}{r^2}\right)^{-1}dr^2+r^2 d\omega_{(3)}^2\ ,
\ee
where $t\in(-\infty,+\infty)$, $r\in(0,\infty)$, and $d\omega^3_{(3)}$ is the metric on a 3-dimensional round sphere, which can be presented in the following form:
\be\n{2.2}
d\omega_{(3)}^2=d\zeta^{2}+\sin^{2}\zeta\,d\vartheta^{2}+\sin^2\zeta\sin^2\vartheta\,d\varphi^2\,, 
\ee
where $\zeta,\vartheta\in[0,\pi]$ and $\varphi\in[0,2\pi)$ are the hyperspherical coordinates. The black hole event horizon is located at $r=r_o$, and the parameter $r_o$ is related to the black hole mass $M$ as follows:
\be\n{M}
r_o^2=\frac{8M}{3\pi}\,.
\ee
The space-time singularity is located at $r=0$.

\begin{figure}[htb]
\begin{center} 
\includegraphics[width=5cm]{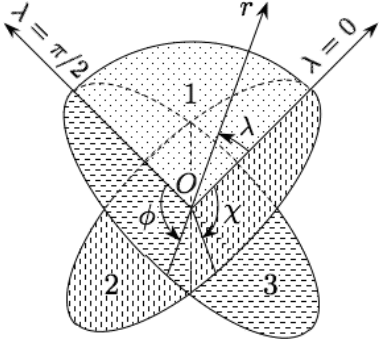} 
\caption{The Hopf coordinates $(\lambda,\chi,\phi)$. The fixed points of the Killing vectors $\xi^\alpha_{(\phi)}$ and $\xi^\alpha_{(\chi)}$ belong to the ``axes" defined by $\lambda=0$ and $\lambda=\pi/2$, respectively. The coordinate origin $O$ is a fixed point of the isometry group $U_\chi(1)\times U_\phi(1)$. Planes 1, 2, and 3, embedded into 4-dimensional space, are orthogonal to each other.} \label{F1} 
\end{center}
\end{figure}

To bring the black hole metric \eq{2.1} to the Weyl form \eq{1.8}, we use the Hopf coordinates $\lambda\in[0,\pi/2]$ and $\chi,\phi\in[0,2\pi)$ in which the metric $d\omega^3_{(3)}$ reads
\be\n{2.3}
d\omega_{(3)}^2=d\lambda^{2}+\cos^2\lambda\,d\chi^{2}+\sin^2\lambda\,d\phi^2\,. 
\ee
Thus, $\chi$ and $\phi$ are Killing coordinates. The space-time \eq{2.1}, \eq{2.3} admits the following orthogonal, commuting Killing vectors:
\be\n{2.4}
\xi^\alpha_{(t)}=\delta^\alpha_{\,\,\, t}\hhh \xi^\alpha_{(\chi)}=\delta^\alpha_{\,\,\, \chi}\hhh \xi^\alpha_{(\phi)}=\delta^\alpha_{\,\,\, \phi}\,,
\ee
where $\xi^\alpha_{(t)}$ is timelike outside the black hole horizon, and $\xi^\alpha_{(\phi)}$, $\xi^\alpha_{(\chi)}$ are spacelike vectors whose fixed points belong to the orthogonal ``axes" $\lambda=0$ and $\lambda=\pi/2$, respectively. The Hopf coordinates are illustrated in Fig.~\ref{F1}. 

It is convenient to introduce the following coordinate transformations:
\ba
r&=&\frac{r_o}{\sqrt{2}}(\eta+1)^{1/2}\hhh \eta\in(-1,\infty),\n{2.5a}\\
\lambda&=&\theta/2\hhh\theta\in[0,\pi]\,.\n{2.5b}
\ea
In the new coordinates $(\eta,\theta)$ the black hole horizon and singularity are located at $\eta=1$ and $\eta=-1$, and the black hole interior and exterior regions are defined by $\eta\in(-1,1)$ and $\eta\in(1,\infty)$, respectively. The metric \eq{2.1}, \eq{2.3} takes the following form: 
\ba\n{2.6}
ds^2&=&-\frac{\eta-1}{\eta+1}dt^2+\frac{r_o^2}{8}(\eta+1)\left[\frac{d\eta^2}{\eta^2-1}+d\theta^2\right.\non\\
&+&\left.2(1+\cos\theta)d\chi^2+2(1-\cos\theta)d\phi^2\right]\,.
\ea
This metric can be written in the Weyl form \eq{1.8} by using the following coordinate transformations:
\ba\n{2.7}
\rho&=&\frac{r_o^2}{4}\sqrt{\eta^2-1}\sin\theta\hhh z=\frac{r_o^2}{4}\eta\cos\theta\,.
\ea
It is more convenient to use $(\eta,\theta)$ coordinates instead of the cylindrical coordinates $(\rho,z)$, which describe the space-time outside the black hole horizon and give additional coordinate singularities in the black hole interior region if analytically continued through the black hole horizon.

The functions $\tu$, $\tw$, and $\tv$ in the coordinates $(\eta,\theta)$ take the following form:
\ba
e^{2\tu}&=&\frac{4}{r_o^2}(\eta+1)^{-1}(1-\cos\theta)^{-1}\,,\n{2.8a}\\
e^{2\tw}&=&\frac{4}{r_o^2}(\eta+1)^{-1}(1+\cos\theta)^{-1}\,,\n{2.9b}\\
e^{2\tv}&=&\frac{r_o^2(\eta+1)^3\sin^2\theta}{8(\eta^2-\cos^2\theta)}\,.\n{2.9c}
\ea
One can check that subject to the transformations \eq{2.7}, the functions $\tu$ and $\tw$ satisfy the Laplace equation \eq{1.0a}, and the function $\tv$ satisfies equations \eq{1.9a} and \eq{1.9b}.
 
\subsection{Metric of a 5-dimensional distorted black hole}

In the previous subsection we demonstrated that the metric of a 5-dimensional Schwarzschild-Tangherlini black hole can be written in the generalized Weyl form \eq{1.8}. Here we present the metric of a 5-dimensional vacuum black hole distorted by external gravitational fields. The fields sources are located at asymptotic infinity and not included into the metric at finite distances. As a result, the corresponding space-time is not asymptotically flat\footnote[2]{Assuming that the external sources are localized at finite distances rather than at infinity, the space-time can be analytically extended to achieve asymptotic flatness in the way described in \cite{Geroch} for a 4-dimensional distorted black hole.}. We consider the space-time near the black hole regular horizon, far away from the sources. In this case, the solution represents a {\em local black hole} in analogy with a 4-dimensional distorted vacuum black hole studied in \cite{Geroch}. We focus on the study of the space-time near the black hole horizon and its interior region, $\eta\in(-1,1)$. The corresponding metric is
\ba
ds^2&=&-\frac{\eta-1}{\eta+1}e^{2(\hu+\hw)}dt^2+\frac{r_o^2}{8}(\eta+1)\left[e^{2(\hv+\hu+\hw)}\right.\non\\
&\times&\left(\frac{d\eta^2}{\eta^2-1}+d\theta^2\right)+2(1+\cos\theta)e^{-2\hw}d\chi^2\non\\
&+&\left.2(1-\cos\theta)e^{-2\hu}d\phi^2\right]\,.\n{2.14}
\ea
In the absence of distortion fields $\hu,\hw$, and $\hv$, this metric reduces to that of the Schwarzschild-Tangherlini black hole \eq{2.6}. The Laplace equation \eq{1.0a} and Eqs. \eq{1.10a} and \eq{1.10b} for the distortion fields $\hu,\hw$, and $\hv$ in the coordinates $(\eta,\theta)$ take the following form:
\ba\n{2.15}
(\eta^2-1)\hx_{,\eta\eta}+2\eta\hx_{,\eta}+\hx_{,\theta\theta}+\cot\theta\hx_{,\theta}=0\,,
\ea
where $\hx:=(\hu,\hw)$, and
\ba
\hv_{,\eta}&=&N\left(\eta\left[(\eta^2-1)(\hu_{,\eta}^2+\hw_{,\eta}^2+\hu_{,\eta}\hw_{,\eta})\right.\right.\non\\
&-&\left.\left.\hu_{,\theta}^2-\hw_{,\theta}^2-\hu_{,\theta}\hw_{,\theta}\right]\right.
+(\eta^2-1)\cot\theta\non\\
&\times&\left[2\hu_{,\eta}\hu_{,\theta}+2\hw_{,\eta}\hw_{,\theta}+\hu_{,\eta}\hw_{,\theta}+\hu_{,\theta}\hw_{,\eta}\right]\non\\
&+&\frac{3}{2}\eta\left[\hu_{,\eta}+\hw_{,\eta}\right]-(\eta^2-1)\frac{\cos\theta}{2\sin^2\theta}\left[\hu_{,\eta}-\hw_{,\eta}\right]\non\\
&+&\left.\frac{3}{2}\cot\theta\left[\hu_{,\theta}+\hw_{,\theta}\right]+\frac{\eta}{2\sin\theta}\left[\hu_{,\theta}-\hw_{,\theta}\right]\right)\non\\
&-&\frac{3}{2}\left[\hu_{,\eta}+\hw_{,\eta}\right]\,,\n{2.16a}\\
\hv_{,\theta}&=&-N\left((\eta^2-1)\cot\theta\left[(\eta^2-1)(\hu_{,\eta}^2+\hw_{,\eta}^2\right.\right.\non\\
&+&\hu_{,\eta}\hw_{,\eta})\left.\left.-\hu_{,\theta}^2-\hw_{,\theta}^2-\hu_{,\theta}\hw_{,\theta}\right]\right.
-\eta(\eta^2-1)\non\\
&\times&\left[2\hu_{,\eta}\hu_{,\theta}+2\hw_{,\eta}\hw_{,\theta}+\hu_{,\eta}\hw_{,\theta}+\hu_{,\theta}\hw_{,\eta}\right]\non\\
&-&\frac{3}{2}\eta\left[\hu_{,\theta}+\hw_{,\theta}\right]+(\eta^2-1)\frac{\cos\theta}{2\sin^2\theta}\left[\hu_{,\theta}-\hw_{,\theta}\right]\non\\
&+&\frac{3}{2}(\eta^2-1)\cot\theta\left[\hu_{,\eta}+\hw_{,\eta}\right]\non\\
&+&\left.\frac{\eta(\eta^2-1)}{2\sin\theta}\left[\hu_{,\eta}-\hw_{,\eta}\right]\right)-\frac{3}{2}\left[\hu_{,\theta}+\hw_{,\theta}\right]\,.\n{2.16b}
\ea
Here $N=\sin^2\theta(\eta^2-\cos^2\theta)^{-1}$ is singular along the lines $\eta=\pm\cos\theta$. However, the function $\hv$, which is given explicitly in the next Section, is regular along these lines. 

If the distortion fields $\hu$ and $\hw$ are known, the function $\hv$ can be derived by the following line integral:
\be\n{2.17}
\hat{V}(\eta,\theta)=\int_{(\eta_0,\theta_0)}^{(\eta,\theta)}\left[\hat{V}_{,\eta'}(\eta',\theta')d\eta'+\hat{V}_{,\theta'}(\eta',\theta')d\theta'\right]\,.
\ee
The integral can be taken along any path connecting the points $(\eta_0,\theta_0)$ and $(\eta,\theta)$. Thus, the field $\hv$ is defined up to arbitrary constant of integration corresponding to the choice of a point $(\eta_0,\theta_0)$. This constant can be chosen to eliminate conical singularities, at least along one connected component of one ``axis''.

Let us note that the distortion fields $\hu$ and $\hw$ define norms of the Killing vectors $\xi^\alpha_{(\phi)}$ and $\xi^\alpha_{(\chi)}$, respectively. Thus, exchange between the ``axes" $\theta=0$ and $\theta=\pi$ is given by the following transformation:
\ba\n{tr}
(\theta,\chi,\phi)&\to&(\pi-\theta,\phi,\chi)\,,\non\\
\\
\left[\hu(\eta,\theta),\hw(\eta,\theta)\right]&\to&\left[\hw(\eta,\theta),\hu(\eta,\theta)\right]\,.\non
\ea
According to Eqs. \eq{2.16a}-\eq{2.17}, the distortion field $\hv$, and hence the metric \eq{2.14}, do not change under this transformation, as it has to be.  

The distorted black hole horizon is defined by $\eta=1$. It is regular, if the space-time invariants are finite on the horizon, and there are no conical singularities along the axes of symmetry, and thus, on the horizon. According to the results presented in Appendix A, the Kretschmann scalar is regular on the black hole horizon if the horizon surface is a regular, totally geodesic surface and its surface gravity is constant. It follows that the distortion fields $\hu$, $\hw$, and $\hv$ must be smooth on a regular horizon. The distortion fields explicitly given in the next Section satisfy this condition.

The metric \eq{2.14} has no conical singularities along the ``axes" $\theta=0$ and $\theta=\pi$, if the space there is locally flat. The no-conical-singularity condition can be formulated as follows: Let us consider a spacelike Killing vector $\xi^\alpha_{(\varphi)}=\delta^\alpha_{\,\,\, \varphi}$, whose orbits are compact near the corresponding symmetry axis defined by $y=y_0$. Let $2\pi$ be the period of the Killing coordinate $\varphi$, and let 
\be\n{2.19}
dl^2=A(y)d\varphi^2+B(y)dy^2\,,
\ee
be a metric of a 2-dimensional surface near the symmetry axis. Then, there is no-conical-singularity corresponding to the symmetry axis if the ratio of the $\xi^\alpha_{(\varphi)}$ orbit circumference at the vicinity of the symmetry axis to the orbit radius, which is defined on the 2-dimensional surface, is equal to $2\pi$, i.e., 
\ba\n{2.20}
\lim_{y\to y_0}\frac{\int_0^{2\pi}A^{1/2}(y)d\varphi}{\int_{y_0}^yB^{1/2}(y')dy'}=\lim_{y\to y_0}\frac{2\pi\lvert A_{,y}(y)\rvert}{2\sqrt{A(y)B(y)}}=2\pi\,.\non\\
\ea
If the ratio is less than $2\pi$ we have angular deficit, and if it is greater than $2\pi$ we have angular excess. 
    
Assuming that the distortion fields $\hu$ and $\hw$ are smooth on the ``axes", the no-conical-singularity condition for the metric \eq{2.14} and for the ``axis" $\theta=0$, where $(x,y)=(\phi,\theta)$, reads
\ba
(\hv+2\hu+\hw)\rvert_{\theta=0}=0\,;\n{2.21a}
\ea 
for the ``axis" $\theta=\pi$, where $(x,y)=(\chi,\theta)$, it is given by
\be
(\hv+\hu+2\hw)\rvert_{\theta=\pi}=0\,.\n{2.21b}
\ee

\section{Solution}

In this Section we derive a solution representing a distorted 5-dimensional vacuum black hole. We start with the Laplace equation \eq{2.15} for the distortion fields $\hu$ and $\hw$. In the cylindrical coordinates $(\rho,z)$ (see, \eq{2.7}) the solution is well known and has the following form:
\ba\n{3.1}
\hx(\rho,z)=\sum_{n\geq0}\left[A_n\,r^{n}+B_n\,r^{-(n+1)}\right]P_n(\cos\vartheta)\,,
\ea
where 
\ba\n{3.2}
r=(\rho^2+z^2)^{1/2}\hhh\cos\vartheta=z/r\,,
\ea
and $P_n(\cos\vartheta)$ are the Legendre polynomials of the first kind. The coefficients $A_n$ and $B_n$ in the expansion \eq{3.1} are called the interior and the exterior multipole moments, respectively (see, e.g., \cite{Multipoles1} and \cite{Multipoles2}). Distortion fields defined by the exterior multipole moments $B_n$'s alone correspond to asymptotically flat solutions. However, according to the uniqueness theorem formulated in \cite{Gary}, a Schwarzschild-Tangherlini black hole is the only $d$-dimensional asymptotically flat static vacuum black hole which has non-degenerate regular event horizon. Note that a combination of the distortion fields corresponding to the exterior and the interior multipole moments makes the black hole horizon ($\rho=0$, $z\in[-r^{2}_{o}/4,r^{2}_{o}/4]$) singular, because the  terms in \eq{3.1} proportional to the $A_n$'s cannot cancel out the divergency at $\rho=z=0$ due to the terms proportional to the $B_n$'s. Thus, to have a regular horizon we shall consider external sources, whose distortion fields are defined by the interior multipole moments $A_n$'s alone.

Applying the coordinate transformations \eq{2.7} to expressions \eq{3.1} and \eq{3.2} we derive  
\ba
&&\hspace{-0.25cm}\hu(\eta,\theta)=\sum_{n\geq0}a_n\,R^{n}P_n(\eta\cos\theta/R)\,,\n{3.3a}\\
&&\hspace{-0.25cm}\hw(\eta,\theta)=\sum_{n\geq0}b_n\,R^{n}P_n(\eta\cos\theta/R)\,,\n{3.3a1}\\
&&\hspace{0.69cm}R=(\eta^2-\sin^2\theta)^{1/2}\,,\n{3.3b}
\ea
where the coefficients $a_n$'s and $b_n$'s define the distortion fields $\hu$ and $\hw$, respectively\footnote[3]{Using the series expansion of the Legendre polynomials (see, e.g., \cite{Arfken}, p. 419)
\be
P_{n}(x)=\frac{1}{2^{n}}\sum^{\lfloor n/2\rfloor}_{k=0}\frac{(-1)^{k}(2n-2k)!}{k!(n-k)!(n-2k)!}x^{n-2k}\,,\non
\ee
where $\lfloor x\rfloor$ is the floor function, one can show that each term $R^{n}P_n(\eta\cos\theta/R)$ in the expansions \eq{3.3a} and \eq{3.3a1} is real valued and regular even when $\eta^{2}\leq\sin^{2}\theta$, which makes $R$ imaginary.}. We shall call these coefficients {\em multipole moments}. In a 4-dimensional space-time, a relation of the multipole moments to their relativistic analogues was discussed in \cite{SuenII}. A general formalism, which includes both the Thorne \cite{Thorne} and the Geroch-Hansen (see, e.g., \cite{Ger1,Ger2,Han,Que}) 4-dimensional relativistic multipole moments is presented in \cite{SuenI}. For a relation between the Thorne \cite{Thorne} and the Geroch-Hansen relativistic multipole moments, see \cite{Beig,Gursel}. 

By analogy with the 4-dimensional case (see, e.g., \cite{Bre1,Bre2}) the distortion field $\hv$ can be presented as a sum of terms linear and quadratic in the multipole moments as follows:   
\ba
&&\hspace{2.2cm}\hv=\hv_1+\hv_2\,,\n{3.4a}\\
&&\hspace{-0.7cm}\hv_1(\eta,\theta)=-\sum_{n\geq0}3(a_n/2+b_n/2)R^nP_n\non\\
&&\hspace{0.45cm}-\,\sum_{n\geq1}\biggl\{(a_n+b_n/2)\sum_{l=0}^{n-1}(\eta-\cos\theta)R^lP_l\non\\
&&\hspace{0.45cm}+\,(a_n/2+b_n)\sum_{l=0}^{n-1}(-1)^{n-l}(\eta+\cos\theta)R^lP_l\biggr\}\,,\n{3.4b}\\
&&\hspace{-0.7cm}\hv_2(\eta,\theta)=\sum_{n,k\geq1}\frac{nk}{n+k}(a_na_k+a_nb_k+b_nb_k)R^{n+k}\non\\
&&\hspace{0.45cm}\times\,[P_nP_k-P_{n-1}P_{k-1}]\,,\, P_n\equiv P_n(\eta\cos\theta/R)\,.\n{3.4c}
\ea
This form of the distortion field $\hv$ corresponds to a particular choice of the constant of integration defined by the initial point $(\eta_0,\theta_0)$ in the line integral \eq{2.17}. Because we have two ``axes", for general $a_n$ and $b_n$ we cannot find such a constant that both the no-conical-singularity conditions \eq{2.21a} and \eq{2.21b} are satisfied simultaneously. To satisfy these conditions we have to impose an additional constraint on the multipole moments $a_n$'s and $b_n$'s. Using the solution \eq{3.3a}--\eq{3.4c}, the no-conical-singularity conditions \eq{2.21a} and \eq{2.21b}, and the symmetry property of the Legendre polynomials 
\be\n{3.5}
P_n(-x)=(-1)^nP_n(x)\,,
\ee
we derive the following constraint on the multipole moments $a_n$'s and $b_n$'s:
\be\n{3.6}
\sum_{n\geq0}(a_{2n}-b_{2n})+3\sum_{n\geq0}(a_{2n+1}+b_{2n+1})=0\,.
\ee
In what follows, we shall refer to the constraint \eq{3.6} as the no-conical-singularity condition for the distorted black hole. One can see that the distortion fields $\hu$, $\hw$, and $\hv$ given by expressions \eq{3.3a}--\eq{3.4c} are smooth on the black hole horizon. Thus, according to the discussion given in the previous Section, the horizon is regular, and this solution represents a local black hole distorted by the external static fields. For this solution the transformation \eq{tr} takes the following form:
\ba\n{tr1}
(\theta,\chi,\phi)&\to&(\pi-\theta,\phi,\chi)\,,\non\\
\\
\left[a_{n},b_{n}\right]&\to&[(-1)^{n}b_{n},(-1)^{n}a_{n}]\,.\non
\ea

An additional restriction on values of the multipole moments follows from the strong energy condition (SEC)
imposed on the external sources of the distortion fields, which follows from the positive mass theorem in a 5-dimensional space-time proven in \cite{Gary2}. If these sources are included, the Einstein equations are not vacuum. In particular, for the metric \eq{2.14} the $\{tt\}$ component of the Einstein equations reads
\ba\n{3.7}
R_{\alpha\beta}\delta^\alpha_{\,\,\, t}\delta^\beta_{\,\,\, t}&=&8\pi\left(T_{\alpha\beta}-\frac{T^\gamma_{\,\,\,\,\gamma}}{3}g_{\alpha\beta}\right)\delta^\alpha_{\,\,\, t}\delta^\beta_{\,\,\, t}\non\\
&=&\frac{\eta-1}{\eta+1}e^{2(\hu+\hw)}\left(\triangle\hu+\triangle\hw\right)\,,
\ea
where $T_{\alpha\beta}$ is the energy-momentum tensor representing the sources. If the sources satisfy SEC, the right hand side of Eq. \eq{3.7} must be non-negative. The Laplace operator $\triangle$ is a negative operator, hence, SEC implies that
\be\n{3.8}
\hu+\hw\leq0\,,
\ee
assuming that $\hu+\hw=0$ at asymptotically flat infinity. In particular, the condition \eq{3.8} implies that on the black hole horizon, on the ``axes" $\theta=0$ and $\theta=\pi$, we have
\be\n{3.9}
\sum_{n\geq0}(\pm1)^n(a_n+b_n)\leq0\,.
\ee
According to the structure of the 5-dimensional Weyl solution, one has an arbitrary choice to define the distortion fields $\hu$ and $\hw$ by specifying the corresponding source functions, which can take any real values (positive or negative), assuming that the SEC \eq{3.9} is satisfied.

To illustrate the effect of the distortion fields on the black hole, we restrict ourselves to the lower order (up to the quadrupole) multipole moments. Values of these moments are subject to the conditions \eq{3.6} and \eq{3.9},
\ba
&&a_0-b_0+a_2-b_2+3(a_1+b_1)=0\,,\n{3.10a}\\
&&a_0+b_{0}\pm (a_1+b_1)+a_{2}+b_2\leq0\,.\n{3.10b}
\ea
The simplest type of distortion is due to a monopole whose values are such that $a_0=b_0\leq0$. However, this distortion is trivial, for it does not break the spherical symmetry of a 5-dimensional Schwarzschild-Tangherlini black hole. The next, less trivial, distortion is due to a dipole. Taking $\hu$ as a dipole distortion and $\hw$ as a monopole distortion and using expression \eq{3.3a}, we derive the dipole-monopole distortion of the form
\ba
\hu&=&a_0+a_1\,\eta\cos\theta\hhh \hw=a_0+3a_1\,,\non\\
&&\hspace{0.7cm}2a_{0}+(3\pm1)a_{1}\leq0\,.\n{3.11}
\ea
According to the transformation \eq{tr1}, taking $\hu$ as a monopole distortion and $\hw$ as a dipole one, corresponds to exchange between the ``axes" $\theta=0$ and $\theta=\pi$ and does not give anything new. Finally, we consider the quadrupole-quadrupole distortion of the form
\ba
\hu&=&\hw=a_0+\frac{a_2}{2}(1-\eta^2+(3\eta^2-1)\cos^2\theta)\,,\non\\
&&\hspace{1.5cm}a_0+a_2\leq0\,.\n{3.12}
\ea   
In what follows, to study the distorted black hole we shall consider the dipole-monopole \eq{3.11} and the quadrupole-quadrupole \eq{3.12} distortion fields. 

\section{Symmetries and boundary values of the distortion fields}

\begin{figure}[htb]
\begin{center} 
\includegraphics[height=5.99cm,width=6cm]{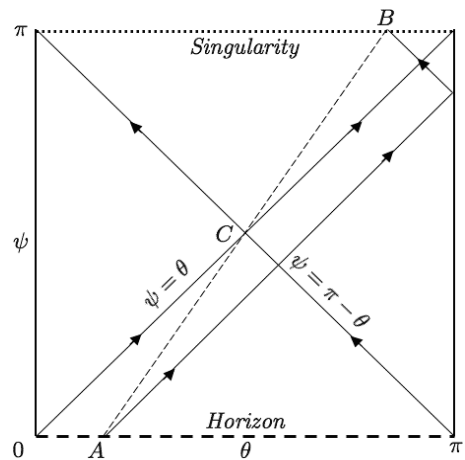} 
\caption{Conformal diagram for the $(\psi$,$\theta$) plane of orbits corresponding to the black hole interior. Arrows illustrate propagation of future directed null rays. Points $A$ and $B$ connected by one of such rays are symmetric with respect to the central point $C(\pi/2, \pi/2)$.} \label{F2} 
\end{center}
\end{figure}

The space-time \eq{2.14} is symmetric under the continuous group of isometries $\mathbb{R}^1_t\times U_\chi(1)\times U_\phi(1)$. This means that the essential features of the space-time geometry are confined to the $(\eta,\theta)$ plane of orbits, which is invariant under the group of transformations. To study the black hole interior, i.e., the region between the black hole horizon and singularity, it is convenient to introduce instead of $\eta$ another coordinate $\psi$ as follows:
\be
\eta=\cos\psi\hhh \psi\in(0,\pi)\,.\n{4.1} 
\ee
Thus, $\psi=0$ and $\psi=\pi$ define the black hole horizon and singularity, respectively. The metric on the plane $(\psi,\theta)$ corresponding to the black hole interior is 
\be
d\Sigma^2=\frac{r_o^2}{8}(1+\cos\psi)e^{2(\hv+\hu+\hw)}\left(-d\psi^2+d\theta^2\right)\,.\n{4.2}
\ee
We see that the coordinate $\psi$ is timelike. The corresponding conformal diagram illustrating the geometry of the black hole interior is presented in Fig.~\ref{F2}. In the diagram, the lines $\psi\pm\theta=const$ are null rays propagating within the 2-dimensional plane $(\psi,\theta)$. Three of such rays are illustrated in Fig.~\ref{F2} by arrows. One of the rays starts at point A on the horizon, goes through the ``axis" $\theta=\pi$, and terminates at the singularity, at point B. 

Consider a transformation $R_C$ representing reflection of a point on the $(\psi,\theta)$ plane with respect to the central point $C$
\be
R_C:\,\,\,(\psi,\theta)~\rightarrow (\pi-\psi,\pi-\theta)\,.\n{4.3}
\ee
This transformation defines a map between functions defined on the plane $(\psi,\theta)$, which has the following form:
\be
f^{*}=R^{*}_C(f):\,\,\, f^*(\psi,\theta)=f(\pi-\psi,\pi-\theta)\,.\n{4.4}
\ee
The coordinates of the points $A$ and $B$ are related by the reflection $R_C$. Thus, $R^{*}_C$ is a map between functions defined on the black hole horizon and singularity. Applying this map to the distortion fields $\hu$, $\hw$, and $\hv$, we derive 
\ba
\hu(\pi-\psi,\pi-\theta)&=&\hu(\psi,\theta)\,,\n{4.5a}\\
\hw(\pi-\psi,\pi-\theta)&=&\hw(\psi,\theta)\,,\n{4.5b}\\
\hv_1(\pi-\psi,\pi-\theta)&=&-\hv_1(\psi,\theta)-3[\hu(\psi,\theta)+\hw(\psi,\theta)]\,,\non\\
\n{4.5c}\\
\hv_2(\pi-\psi,\pi-\theta)&=&\hv_2(\psi,\theta)\,.\n{4.5d}
\ea
We shall use these relations to define values of the distortion fields on the black hole horizon and singularity, as well as on the symmetry ``axes". 

To begin with let us introduce the following notations:
\ba
u_0&:=&\sum_{n\geq0}a_{2n}\hhh u_1:=\sum_{n\geq0}a_{2n+1}\,,\n{4.6a}\\
w_0&:=&\sum_{n\geq0}b_{2n}\hhh w_1:=-\sum_{n\geq0}b_{2n+1}\,,\n{4.6b}
\ea
Then the no-conical-singularity condition \eq{3.6} can be written as
\be\n{4.7}
u_0+3u_1=w_0+3w_1\,.
\ee
In addition, we define the following functions:
\ba
u_{\pm}(\sigma)&:=&\sum_{n\geq0}(\pm1)^na_{n}\cos^n(\sigma)-u_0\,,\n{4.8a}\\
w_{\pm}(\sigma)&:=&\sum_{n\geq0}(\pm1)^nb_{n}\cos^n(\sigma)-w_0\,,\n{4.8b}
\ea
where $\sigma:=(\psi,\theta)$. Thus, for the dipole-monopole distortion \eq{3.11} we have
\ba
u_{\pm}(\sigma)&=&\pm a_1\cos\sigma\hhh u_0=a_0\hhh u_1=a_1\,,\non\\
w_{\pm}(\sigma)&=&0\hhh w_0=a_0+3a_1\hhh w_1=0\,,\n{3.11a}
\ea
and for the quadrupole-quadrupole distortion \eq{3.12} we have
\ba
u_{\pm}(\sigma)&=&w_{\pm}(\sigma)=-a_2\sin^2\sigma\,,\non\\
u_0&=&w_0=a_0+a_2\hhh u_1=w_1=0\,.\n{3.12a}
\ea

Using the definitions above it is convenient to introduce renormalized distortion fields, which do not depend on the monopole moments $a_{0}$ and $b_{0}$, as follows:
\ba
\cu(\psi,\theta)&:=&\hu(\psi,\theta)-u_0-3u_1\,,\n{4.9a}\\
\cw(\psi,\theta)&:=&\hw(\psi,\theta)-w_0-3w_1\,,\n{4.9b}\\
\cv(\psi,\theta)&:=&\hv(\psi,\theta)+\frac{3}{2}\left[u_0+w_0+3(u_1+w_1)\right].\n{4.9c}
\ea
With the aid of the expressions above we derive values of the renormalized distortion fields on the black hole horizon
\ba
\cu(0,\theta)&:=&u_+(\theta)-3u_1\,,\n{4.10a}\\
\cw(0,\theta)&:=&w_+(\theta)-3w_1\,,\n{4.10b}\\
\cv(0,\theta)&:=&4(u_1+w_1)\,,\n{4.10c}
\ea
and the singularity
\ba
&&\hspace{-0.5cm}\cu(\pi,\theta)=u_-(\theta)-3u_1\,,\n{4.11a}\\
&&\hspace{-0.5cm}\cw(\pi,\theta)=w_-(\theta)-3w_1\,,\n{4.11b}\\
&&\hspace{-0.5cm}\cv(\pi,\theta)=-3(u_-(\theta)+w_-(\theta))+5(u_1+w_1)\,,\n{4.11c}
\ea
as well as on the ``axis" $\theta=0$
\ba
\cu(\psi,0)&=&u_+(\psi)-3u_1\,,\n{4.12a}\\
\cw(\psi,0)&=&w_+(\psi)-3w_1\,,\n{4.12b}\\
\cv(\psi,0)&=&-2u_+(\psi)-w_+(\psi)+3(2u_1+w_1)\,,\n{4.12c}
\ea
and on the ``axis" $\theta=\pi$
\ba
\cu(\psi,\pi)&=&u_-(\psi)-3u_1\,,\n{4.12d}\\
\cw(\psi,\pi)&=&w_-(\psi)-3w_1\,,\n{4.12e}\\
\cv(\psi,\pi)&=&-u_-(\psi)-2w_-(\psi)+3(u_1+2w_1)\,.\n{4.12f}
\ea

In what follows, we consider for convenience the dimensionless form of the metric $dS^2$, which is related to the metric $ds^2$ as follows:
\be\n{4.13}
ds^2=\Omega^2dS^2,
\ee
where 
\be\n{4.13a}
\Omega^2=\frac{1}{\kappa^2}e^{2(u_0-u_1+w_0-w_1)}\,
\ee
is the conformal factor, and $\kappa$ is the surface gravity of the distorted black hole corresponding to $\xi^\alpha_{(t)}=\delta^\alpha_{\,\,\, t}$,
\be\n{4.14}
\kappa=\left.\frac{e^{-\hv}}{r_0}\right|_{\eta=1}=\frac{1}{r_0}e^{(3u_0+u_1+3w_0+w_1)/2}\,.
\ee
Note that the space-time \eq{2.14} is not asymptotically flat, so the surface gravity \eq{4.14} is defined only up to an arbitrary red-shift factor. The dimensionless metric is given by
\ba\n{4.15}
dS^2&=&-\frac{\eta-1}{\eta+1}e^{2(\cu+\cw)}dT^2+\frac{1}{8}(\eta+1)\left[e^{2(\cv+\cu+\cw)}\right.\non\\
&\times&\left(\frac{d\eta^2}{\eta^2-1}+d\theta^2\right)+2(1+\cos\theta)e^{-2\cw}d\chi^2\non\\
&+&\left.2(1-\cos\theta)e^{-2\cu}d\phi^2\right]\,,
\ea
where the dimensionless time $T$ is defined as follows: 
\be\n{4.16}
T=\kappa\,e^{4(u_1+w_1)}t\,.
\ee
Using the transformation \eq{4.1}, one can present the metric \eq{4.15} in $(T,\psi,\theta,\chi,\phi)$ coordinates, which are more convenient for analysis of the black hole interior. 

\section{Space-time near the horizon}

\subsection{Intrinsic curvature of the horizon surface}

In this Section we study geometry of the 3-dimensional distorted horizon surface of the space-time \eq{4.15}, defined by $T=const$, $\eta=1$. The metric of the horizon surface reads
\ba
d\Sigma^2_{+}&=&\frac{1}{4}\biggl(e^{2(u_+(\theta)+w_+(\theta)+u_1+w_1)}d\theta^2\non\\
&+&2(1+\cos\theta)e^{-2(w_+(\theta)-3w_1)}d\chi^2\non\\
&+&2(1-\cos\theta)e^{-2(u_+(\theta)-3u_1)}d\phi^2\biggr)\,.\n{5.1}
\ea
Here and in what follows, the $`+'$ subscript stands for a quantity defined on the black hole horizon surface. Using this metric one can calculate the dimensionless area of the black hole horizon surface,
\be\n{5.2}
\mathcal{A}_+=2\pi^2e^{4(u_1+w_1)}\,.
\ee
The dimensional area is equal to 
\be\n{dar}
A_{+}=\Omega^3\mathcal{A}_+=2\pi^{2}r_{o}^{3}e^{-(u_{1}+w_{1}+3u_{0}+3w_{0})/2}\,.
\ee
To study the geometry of a 2-dimensional surface, one can calculate its intrinsic (Gaussian) curvature invariant and illustrate its shape by an isometric embedding of the surface into a 3-dimensional flat space; one can calculate its extrinsic curvature as well. To study the geometry of a 3-dimensional hypersurface is not that simple, for there are generally more than one curvature invariant, and its isometric local embedding generally requires $3(3+1)/2=6$-dimensional flat space. However, if the hypersurface admits a group of isometries, one can analyze its geometry by studying the geometry of the sections of the isometry orbits. In our case the 3-dimensional hypersurface defined by the metric \eq{5.1} admits a $U_\chi(1)\times U_\phi(1)$ group of isometries. As a result, we have $(\theta,\chi)$ and $(\theta,\phi)$ 2-dimensional sections. For completeness, we consider $(\chi,\phi)$ 2-dimensional sections as well. Following an analysis of the horizon surface of a 5-dimensional black hole and black ring presented in \cite{FR}, we define the Gaussian curvatures of the sections as the corresponding Riemann tensor components of the metric \eq{5.1} calculated in an orthonormal frame,
\ba
K_{+\phi}&:=&\frac{8(1-\cos\theta)}{\sin^{2}\theta}e^{-2(u_+(\theta)+u_1+4w_1)}{\cal R}_{+\theta\chi\theta\chi}\,,\n{K1}\\
K_{+\chi}&:=&\frac{8(1+\cos\theta)}{\sin^{2}\theta}e^{-2(w_+(\theta)+w_1+4u_1)}{\cal R}_{+\theta\phi\theta\phi}\,,\n{K2}\\
K_{+\theta}&:=&\frac{4}{\sin^{2}\theta}e^{2(u_+(\theta)+w_{+}(\theta)-3u_1-3w_1)}{\cal R}_{+\chi\phi\chi\phi}\,.\n{K3}
\ea
Explicit form of these expressions is presented in Appendix B. For a round 3-dimensional sphere, which represents the horizon surface of a 5-dimensional Schwarzschild-Tangherlini black hole, we have
\be
K_{+\phi}=K_{+\chi}=K_{+\theta}=1.
\ee
In the case of the distortion fields $\hu=0$, $\hw\ne0$ we have $K_{+\chi}=K_{+\theta}$, and in the case of the distortion fields $\hu\ne0$, $\hw=0$ we have $K_{+\phi}=K_{+\theta}$. 
\begin{figure*}[htb]
\begin{center}
\ba
&&\hspace{0cm}\includegraphics[width=6cm]{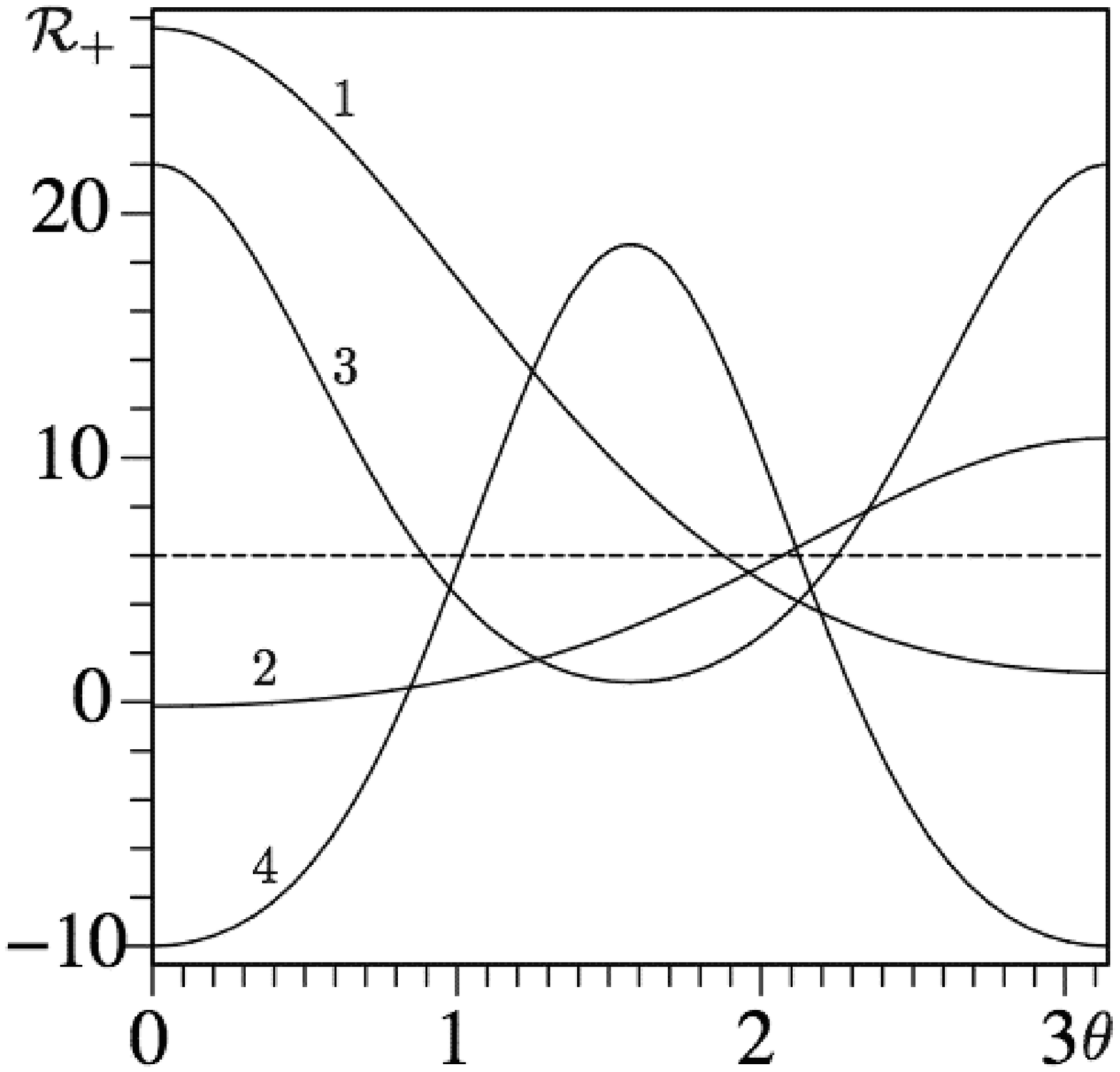}
\hspace{2cm}\includegraphics[width=7.4cm]{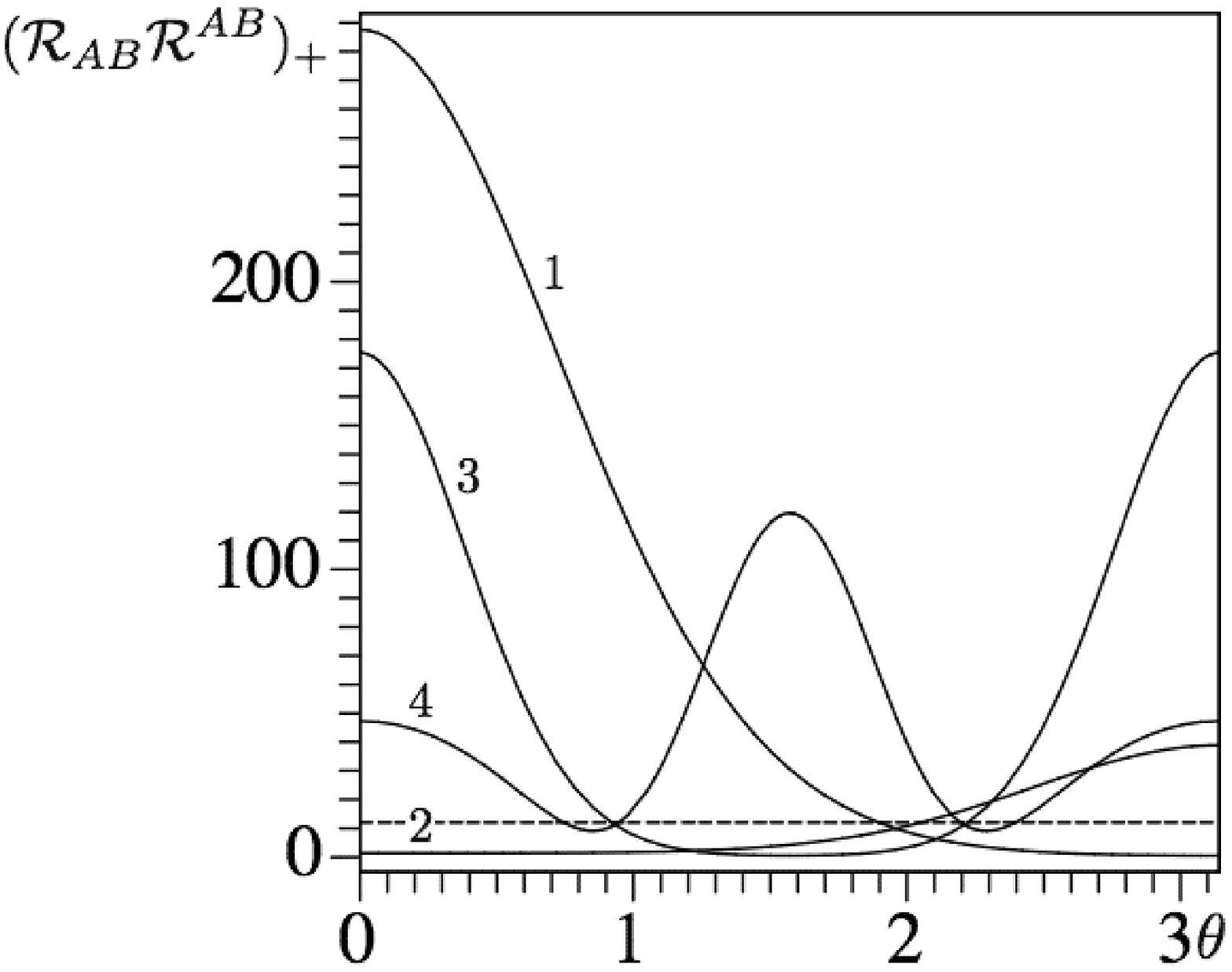}\non\\ 
&&\hspace{3.1cm}({\bf a})\hspace{9cm}({\bf b})\non
\ea
\caption{Intrinsic curvature invariants of the horizon surface. ({\bf a}) Dimensionless Ricci scalar. ({\bf b}) The trace of the square of the Ricci tensor. Dipole-monopole distortion: $a_1=-1/5$, $b_{1}=0$ (line 1), $a_1=1/5$, $b_{1}=0$ (line 2). Quadrupole-quadrupole distortion: $a_2=b_{2}=-1/7$ (line 3), $a_2=b_{2}=1/7$ (line 4). The horizontal dashed lines represent the dimensionless Ricci scalar and the trace of the square of the Ricci tensor of a Schwarzschild-Tangherlini black hole.}\label{F3}
\end{center} 
\end{figure*} 
Components of the Ricci tensor corresponding to a 3-dimensional hypersurface are related to the Gaussian curvatures of the sections as follows:
\ba\n{KR} 
&&{\cal R}^{\,\phi}_{+\phi}=K_{+\theta}+K_{+\chi}\hhh{\cal R}^{\,\chi}_{+\chi}=K_{+\theta}+K_{+\phi}\,,\non\\
&&\hspace{1.5cm}{\cal R}^{\,\theta}_{+\theta}=K_{+\chi}+K_{+\phi}\,.
\ea
The corresponding Ricci scalar and the trace of the square of the Ricci tensor are
\ba
{\cal R}_{+}&=&{\cal R}^{\,\phi}_{+\phi}+{\cal R}^{\,\chi}_{+\chi}+{\cal R}^{\,\theta}_{+\theta}\,,\n{B2}\\
({\cal R}_{AB}{\cal R}^{AB})_{+}&=&({\cal R}^{\,\phi}_{+\phi})^2+({\cal R}^{\,\chi}_{+\chi})^{2}+({\cal R}^{\,\theta}_{+\theta})^2\,.\n{B3}
\ea
The Ricci scalar ${\cal R}_+$ and the trace of the square of the Ricci tensor, $({\cal R}_{AB}{\cal R}^{AB})_+$, of the horizon surface are natural invariant measures of its intrinsic curvature. The dimensional Ricci scalar and the trace of the square of the Ricci tensor are equal to $\Omega^{-2}{\cal R}_+$ and $\Omega^{-4}({\cal R}_{AB}{\cal R}^{AB})_+$, respectively. 

Here we calculate the Gaussian curvatures of the sections for the dipole-monopole distortion \eq{3.11a},
\ba
&&K_{+\phi}=K_{+\theta}=e^{-2a_1(1+\cos\theta)}\left[1+2a_1(1-\cos\theta)\right]\,,\n{5.3a}\\
&&K_{+\chi}=e^{-2a_1(1+\cos\theta)}\left[1-2a_1(3+5\cos\theta)-8a^{2}_{1}\sin^{2}\theta\right]\,,\non\\
\n{5.3b}
\ea
and for the quadrupole-quadrupole distortion \eq{3.12a},
\ba
&&\hspace{2.0cm}K_{+\phi}=k_{+}\hhh K_{+\chi}=k_{-}\,,\non\\
&&\hspace{0.5cm}k_{\pm}=e^{4a_{2}\sin^{2}\theta}\left[1+8a_{2}(1\pm2\cos\theta-4\cos^{2}\theta)\right.\non\\
&&\hspace{1.0cm}-\left.48a^{2}_{2}\cos^{2}\theta\sin^{2}\theta\right]\n{5.4a}\\
&&\hspace{0.0cm}K_{+\theta}=e^{4a_{2}\sin^{2}\theta}\left[1-8a_{2}\cos^{2}\theta-16a^{2}_{2}\cos^{2}\theta\sin^{2}\theta\right]\,.\non\\
\n{5.4b}
\ea
Using these expressions together with Eqs. \eq{KR}--\eq{B3} we can calculate the corresponding dimensionless Ricci scalar and the trace of the square of the Ricci tensor of the horizon surface. For an undistorted black hole the dimensionless Ricci scalar is ${\cal R}_{{\ind{ST}}_{+}}=6$, and the trace of the square of the Ricci tensor is $({\cal R}_{AB}{\cal R}^{AB})_{{\ind{ST}}_{+}}=12$. The Ricci scalar and the trace of the square of the Ricci tensor are shown in Figs.~\ref{F3}({\bf a}) and ({\bf b}), respectively. These figures illustrate that the intrinsic curvature of a distorted horizon surface strongly varies over it.

\begin{figure*}[htb]
\begin{center}
\ba
&&\hspace{0cm}\includegraphics[width=5cm]{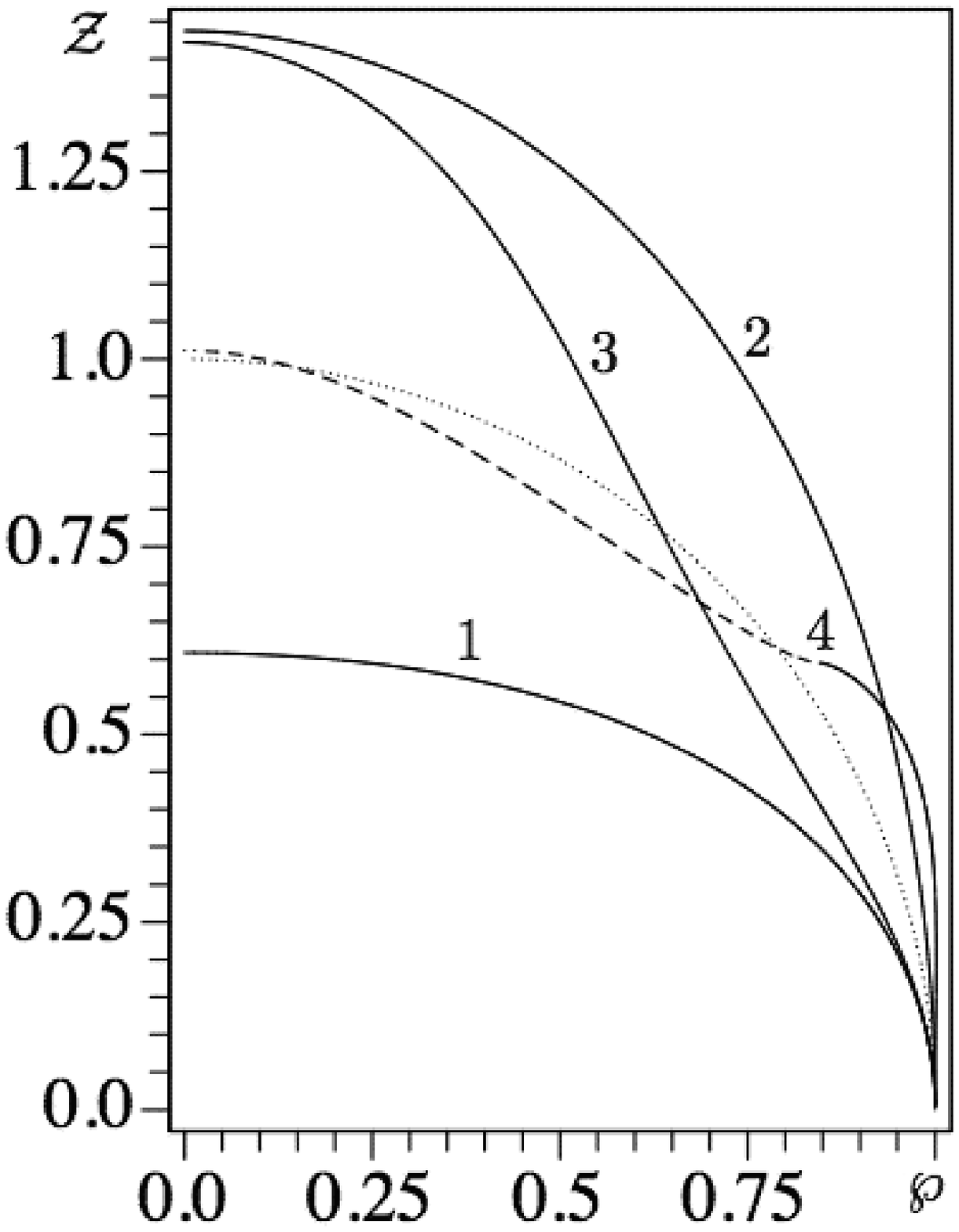}
\hspace{2cm}\includegraphics[width=8cm]{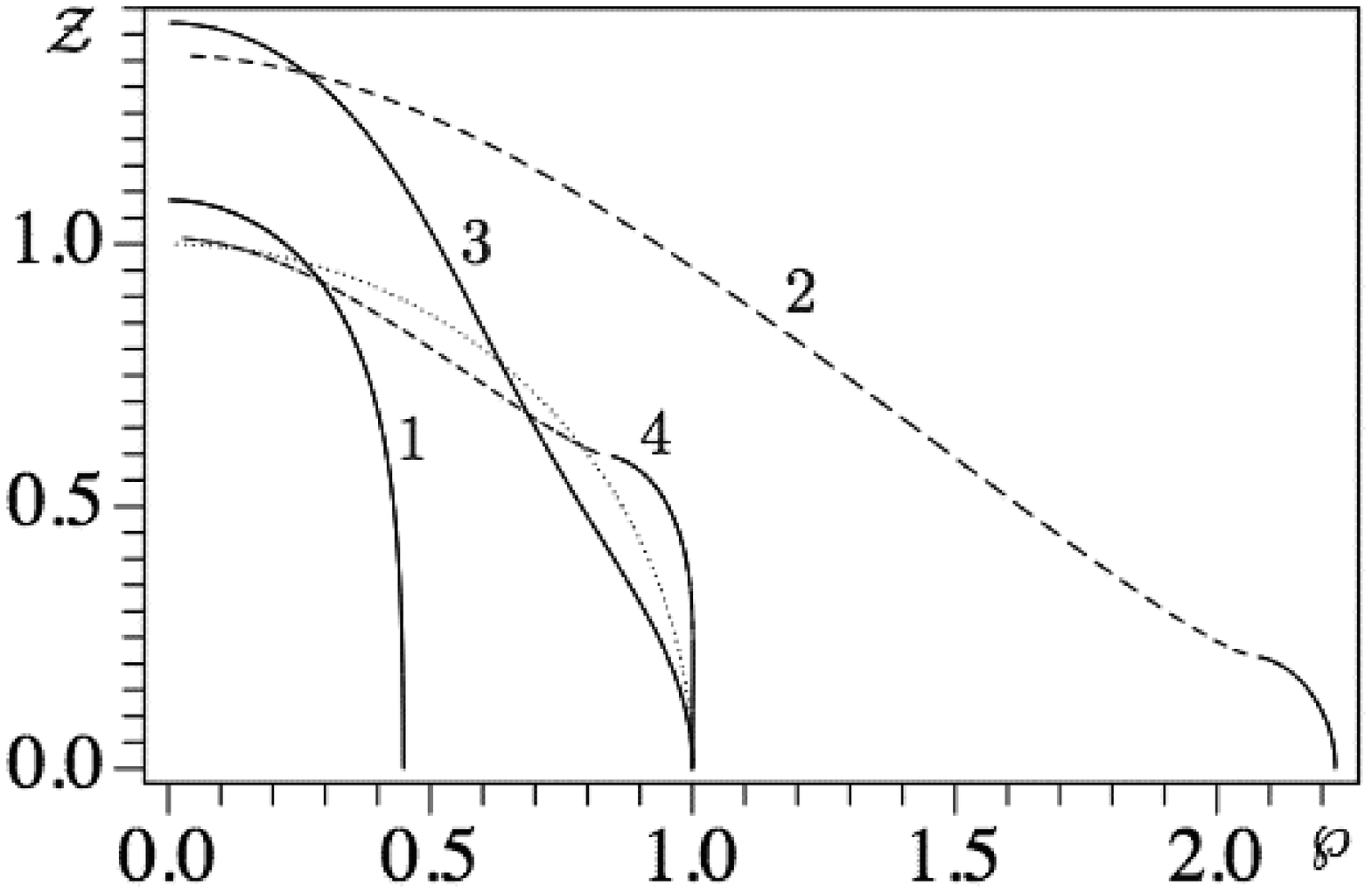}\non\\ 
&&\hspace{2.6cm}({\bf a})\hspace{8.1cm}({\bf b})\non
\ea
\caption{Rotational curves of the horizon surface. ({\bf a}): Section $(\theta,\chi)$. ({\bf b}): Section $(\theta,\phi)$.  Dipole-monopole distortion: $a_1=-1/5$, $b_{1}=0$ (line 1), $a_1=1/5$, $b_{1}=0$ (line 2). Quadrupole-quadrupole distortion: $a_2=b_{2}=-1/7$ (line 3), $a_2=b_{2}=1/7$ (line 4). Regions of the sections embedded into pseudo-Euclidean space are illustrated by dashed lines. Dotted arcs of unit radius represent the horizon surface of an undistorted black hole.}\label{F4}
\end{center} 
\end{figure*}

\subsection{Shape of the horizon surface}

Distortion fields change the shape of the horizon surface. To visualize the effect of the distortion fields on the horizon surface, we consider an isometric embedding of its 2-dimensional sections into a flat 3-dimensional space with the following metric: 
\be\n{5.3}
dl^2=\epsilon d{\cal Z}^2+d\wp^2+\wp^2d\varphi^2,
\ee
where $\epsilon=+1$ corresponds to Euclidean space, $\epsilon=-1$ corresponds to pseudo-Euclidean space, and $({\cal Z},\wp,\varphi)$ are the cylindrical coordinates. 

The section $(\chi,\phi)$ defined by $\theta=const$ represents a 2-dimensional torus whose radii are defined by the distortion fields. We shall consider the embedding of the $(\theta,\chi)$ and $(\theta,\phi)$ 2-dimensional sections, which according to the metric \eq{5.1} are parametrized in the cylindrical coordinates as follows:
\be\n{5.4}
{\cal Z}={\cal Z}(\theta)\hhh \wp=\wp(\theta)\,.
\ee
  
The geometry induced on the section \eq{5.4} is given by
\be\n{5.5}
dl^2=(\epsilon {\cal Z}^2_{,\theta}+\wp^2_{,\theta})d\theta^2+\wp^2d\varphi^2\,.
\ee

The metric of the section $(\theta,\chi)$ defined from the metric \eq{5.1} by $\phi=const$ reads 
\ba\n{5.6}
d\Sigma^2_{+\phi}&=&\frac{1}{4}\biggl(e^{2(u_{+}(\theta)+w_+(\theta)+u_1+w_1)}d\theta^2\non\\
&+&2(1+\cos\theta)e^{-2(w_{+}(\theta)-3w_1)}d\chi^2\biggr)\,.
\ea 
Matching the metrics \eq{5.5} and \eq{5.6}, we derive the embedding map
\ba\n{5.7}
&&\hspace{0cm}\varphi=\chi\hhh \wp(\theta)=\frac{1}{\sqrt{2}}(1+\cos\theta)^{1/2}e^{-w_{+}(\theta)+3w_1}\,,\non\\
&&\hspace{1.7cm}{\cal Z}(\theta)=\int_0^{\theta}{\cal Z}_{,\theta'}d\theta'\,,\non\\
&&\hspace{-0.5cm}{\cal Z}_{,\theta}=\left[\epsilon\left(\frac{1}{4}e^{2(u_{+}(\theta)+w_{+}(\theta)+u_{1}+w_1)}-\wp^{2}_{,\theta}\right)\right]^{1/2}\,.
\ea

The metric of the section $(\theta,\phi)$ defined from the metric \eq{5.1} by $\chi=const$ reads 
\ba\n{5.8}
d\Sigma^2_{+\chi}&=&\frac{1}{4}\biggl(e^{2(u_{+}(\theta)+w_+(\theta)+u_1+w_1)}d\theta^2\non\\
&+&2(1-\cos\theta)e^{-2(u_{+}(\theta)-3u_1)}d\phi^2\biggr)\,.
\ea
Matching the metrics \eq{5.5} and \eq{5.8}, we derive the embedding map
\ba\n{5.9}
&&\hspace{0cm}\varphi=\phi\hhh \wp(\theta)=\frac{1}{\sqrt{2}}(1-\cos\theta)^{1/2}e^{-u_{+}(\theta)+3u_1}\,,\non\\
&&\hspace{1.7cm}{\cal Z}(\theta)=\int_{\pi}^{\theta}{\cal Z}_{,\theta'}d\theta'\,,\non\\
&&\hspace{-0.5cm}{\cal Z}_{,\theta}=-\left[\epsilon\left(\frac{1}{4}e^{2(u_{+}(\theta)+w_{+}(\theta)+u_{1}+w_1)}-\wp^{2}_{,\theta}\right)\right]^{1/2}\,.
\ea

Rotational curves illustrating embeddings of the sections $(\theta,\chi)$ and $(\theta,\phi)$ for the dipole-monopole \eq{3.11a} and the quadrupole-quadrupole \eq{3.12a} distortions are shown in Figs.~\ref{F4}({\bf a}) and \ref{F4}({\bf b}), respectively. These curves belong to plane 1 in Fig.~\ref{F1}. To reconstruct the shape of the 3-dimensional horizon surface, we have to rotate these curves in planes 2 and 3 (see Fig.~\ref{F1}) around the ``axes" $\lambda=\pi/2$ and $\lambda=0$.

\subsection{Metric near the horizon}

The functions $u_+(\theta)$ and $w_+(\theta)$, which specify the geometry of the horizon surface, uniquely determine the space-time geometry in the vicinity of the black hole horizon. Using the expansion of the distortion fields $\hu$, $\hw$, and $\hv$ in the vicinity of the horizon (see Eqs. \eq{C7a} and \eq{C9a} in Appendix C) and the definition of the renormalized distortion fields \eq{4.9a}--\eq{4.9c}, we derive an approximation for the metric \eq{4.15} near the black hole horizon:
\be\n{5.10}
dS_+^{2}=A_+ dT^2+B_+(-d\psi^2+d\theta^2)+C_+d\chi^2+D_+d\phi^2\, ,
\ee
\ba
A_+&=&\frac{1}{4}e^{2(u_+(\theta)+w_+(\theta)-3u_1-3w_1)}\biggl[\psi^2+\frac{1}{2}\biggl(u_{+,\theta\theta}+w_{+,\theta\theta}\non\\
&+&\left.\left.\cot\theta(u_{+,\theta}+w_{+,\theta})+\frac{1}{3}\right)\psi^4+{\cal O}(\psi^{6})\right]\ ,\non\\
B_+&=&\frac{1}{4}e^{2(u_+(\theta)+w_+(\theta)+u_1+w_1)}\biggl[1+\frac{1}{2}\biggl(u_{+,\theta\theta}+w_{+,\theta\theta}\non\\
&+&2(u_{+,\theta}^2+u_{+,\theta}w_{+,\theta}+w_{+,\theta}^2)-2\cot\theta(u_{+,\theta}+w_{+,\theta})\non\\
&-&\left.\left.\frac{u_{+,\theta}-w_{+,\theta}}{\sin\theta}-\frac{1}{2}\right)\psi^2+{\cal O}(\psi^{4})\right]\,,\non\\
C_+&=&\frac{1}{2}(1+\cos\theta)e^{-2(w_+(\theta)-3w_1)}\non\\
&\times&\left[1-\frac{1}{2}\left(w_{+,\theta\theta}+\cot\theta\,w_{+,\theta}+\frac{1}{2}\right)\psi^2+{\cal O}(\psi^{4})\right]\,,\non\\
D_+&=&\frac{1}{2}(1-\cos\theta)e^{-2(u_+(\theta)-3u_1)}\non\\
&\times&\left[1-\frac{1}{2}\left(u_{+,\theta\theta}+\cot\theta\,u_{+,\theta}+\frac{1}{2}\right)\psi^2+{\cal O}(\psi^{4})\right]\,.\non\\
\n{5.11}
\ea
This approximation allows us to calculate the Kretschmann scalar ${\cal K}:=^{(5)}\hspace{-0.4cm}R_{\alpha\beta\gamma\delta}\hspace{0.05cm}^{(5)}\hspace{-0.09cm}R^{\alpha\beta\gamma\delta}$, which is a space-time curvature invariant, at the horizon surface. In Appendix A we demonstrate that there is a simple relation between the Kretschmann scalar calculated on the horizon of a 5-dimensional, static, distorted black hole and the trace of the square of the Ricci tensor of its horizon surface, which is
\be\n{5.12}
{\cal K}_+=6({\cal R}_{AB}{\cal R}^{AB})_+\,.
\ee
This relation is valid not only for a distorted black hole given by a 5-dimensional Weyl solution, but also for an arbitrary distorted, asymmetric, static, vacuum 5-dimensional black hole.

Consequently, according to figures \ref{F3} and \ref{F4}, the space-time curvature at the horizon is greater at the points where the horizon surface is more curved.

\section{Space-time near the singularity}

\subsection{Metric near the singularity}

Using the expansion of the distortion fields $\hu$, $\hw$, and $\hv$ at the vicinity of the black hole singularity (see Eqs. \eq{C7a} and \eq{C9b} in Appendix C) and the definition of the renormalized distortion fields \eq{4.9a}--\eq{4.9c}, we derive an approximation of the metric \eq{4.15} near the black hole singularity $\psi_-=\pi-\psi \to 0:$
\be
dS_-^{2}=A_- dT^2+B_-(-d\psi_-^2+d\theta^2)+C_-d\chi^2+D_-d\phi^2\,,\n{6.1}
\ee
\ba
A_-&=&4e^{2(u_-(\theta)+w_-(\theta)-3u_1-3w_1)}\biggl[\frac{1}{\psi_-^2}+\frac{1}{2}\biggl(u_{-,\theta\theta}\non\\
&+&w_{-,\theta\theta}\left.\left.+\cot\theta(u_{-,\theta}+w_{-,\theta})-\frac{1}{3}\right)+{\cal O}(\psi_-^{2})\right]\ ,\non\\
B_-&=&\frac{1}{16}e^{-4(u_-(\theta)+w_-(\theta)-u_1-w_1)}\biggl[\psi_-^2-\biggl(u_{-,\theta\theta}+w_{-,\theta\theta}\non\\
&-&u_{-,\theta}^2-u_{-,\theta}w_{-,\theta}-w_{-,\theta}^2-\frac{1}{2}\cot\theta(u_{-,\theta}+w_{-,\theta})\non\\
&+&\left.\left.\frac{u_{-,\theta}-w_{-,\theta}}{2\sin\theta}+\frac{1}{12}\right)\psi_-^4+{\cal O}(\psi_-^{6})\right]\,,\non\\
C_-&=&\frac{1}{8}(1+\cos\theta)e^{-2(w_-(\theta)-3w_1)}\biggl[\psi_-^2\non\\
&-&\left.\frac{1}{2}\left(w_{-,\theta\theta}+\cot\theta\,w_{-,\theta}+\frac{1}{6}\right)\psi_-^4+{\cal O}(\psi_-^{6})\right]\,,\non\\
D_-&=&\frac{1}{8}(1-\cos\theta)e^{-2(u_-(\theta)-3u_1)}\biggl[\psi_-^2\non\\
&-&\left.\frac{1}{2}\left(u_{-,\theta\theta}+\cot\theta\,u_{-,\theta}+\frac{1}{6}\right)\psi_-^4+{\cal O}(\psi_-^{6})\right]\,.\non\\
\n{6.2}
\ea
This approximation allows us to calculate the Kretschmann scalar near the singularity, up to corrections that are second order in $\psi_-$:
\ba
&&\hspace{-0.3cm}{\cal K}_-\approx\frac{2^8\cdot72}{\psi_-^8}e^{8(u_-(\theta)+w_-(\theta)-u_1-w_1)}\left[1+ {\cal K}_-^{(2)}\psi_-^2\right]\,,\n{6.3a}\\
&&\hspace{-0.3cm}{\cal K}_-^{(2)}=\frac{2}{3}\biggl(u_{-,\theta\theta}+w_{-,\theta\theta}-4u_{-,\theta}^2-6u_{-,\theta}w_{-,\theta}-4w_{-,\theta}^2\biggr.\non\\
&&\hspace{0.3cm}-\left.2\cot\theta(u_{-,\theta}+w_{-,\theta})+\frac{u_{-,\theta}-w_{-,\theta}}{\sin\theta}+\frac{1}{2}\right)\,.\n{6.3b}
\ea
Higher order terms can be obtained by using the relations given in Appendix C. In the absence of distortion the Kretschmann scalar is equal to the Kretschmann scalar of the 5-dimensional Schwarzschild-Tangherlini space-time
\be\n{6.4}
{\cal K}_{{\ind{ST}}_{-}}=\frac{2^8\cdot72}{\psi_-^{8}}\,.
\ee

\subsection{Stretched singularity}

In the absence of distortion the approximation \eq{6.2} gives the Schwarzschild-Tangherlini geometry near the singularity
\be\n{6.5}
dS_{-}^2\approx -\frac{\psi_-^2}{16}d\psi_-^2+\frac{4}{\psi_-^2} dT^2+\frac{\psi_-^2}{4}d\omega_{(3)}^2\,.
\ee
Using the transformation
\be\n{6.6}
\psi_-=2\sqrt{2}\tau^{1/2}
\ee
the metric \eq{6.5} can be written in the form
\be\n{6.7}
dS_{-}^2\approx -d\tau^2+\frac{1}{2\tau}\,dT^2+2\tau\,d\omega_{(3)}^2\,.
\ee
Here $\tau$ is the maximal proper time of free fall to the singularity from a point near it along the geodesic defined by $(T,\theta,\chi,\phi)=const$. The proper time $\tau$ is positive and equals to $0$ at the singularity\footnote[4]{The proper time $\tau$ defined this way runs backward. One can define another proper time $\tau':=\tau_o-\tau$, where $\tau_o\geq\tau$, which runs forward and is equal to $\tau_o$ at the singularity. However, we shall use the former definition, which is more convenient for our calculations.}. The metric \eq{6.7} has the Kasner exponents $(-1/2,1/2,1/2,1/2)$. It represents a metric of a collapsing, anisotropic universe which contracts in the ($\theta$,$\chi$,$\phi$)-directions and expands in the $T$-direction.

The Kretschmann scalar \eq{6.4}, expressed through the proper time, has the following form:
\be\n{6.8}
{\cal K}_{{\ind{ST}}_{-}}=\frac{9}{2\tau^4}\,.
\ee
This expression shows that a surface of constant ${\cal K}_{{\ind{ST}}_{-}}$ is at the same time a surface of
constant $\tau$.

A space-time in the region where its curvature is of order of the Planckian curvature requires quantum gravity for its description. For the Schwarzschild-Tangherlini geometry such a region is defined by the surface where ${\cal K}_{{\ind{ST}}_{-}}\sim\ell_{\ind{Pl}}^{-4}$, where $\ell_{\ind{Pl}}\sim10^{-33}$cm is the Planckian length, which corresponds to the proper time $\tau$ of order of the Planckian time $\tau_{\ind{Pl}}\sim10^{-44}$s.  Since one cannot rely on the classical description in this region, it is natural to consider its boundary as the {\em stretched
singularity}. The stretched singularity of the 5-dimensional Schwarzschild-Tangherlini space-time has the topology $\mathbb{R}^1\times S^3$. Its metric is a direct sum of the metric of a line and the metric of a round 3-dimensional sphere.

What happens to the stretched singularity when a Schwarzschild-Tangherlini black hole is distorted? To answer this question we use the asymptotic form of the metric near the singularity of the distorted black hole (see Eq. \eq{6.1}). Let us consider a timelike geodesic defined by $(T,\chi,\phi)=const$. For such a geodesic the maximal proper time of free fall to the singularity from a point near it corresponds to $E=L_\chi=L_\phi=L_{0}=0$
(see Appendix D). We shall call the corresponding geodesic ``radial''. According to the calculations given in Appendix D, the ``radial'' geodesic is uniquely determined by the limiting value $\theta_{0}$ of its angular parameter $\theta$ at which it ``hits'' the black hole singularity. Let us denote by $\tau$ the proper time measured along the ``radial'' geodesic backward in time from its end point at the singularity. We can use $(\tau,\theta_{0})$ as new coordinates in the vicinity of the singularity. Using the leading order terms in expressions \eq{D11a} and \eq{D11b}, we can relate the coordinates $(\psi_-,\theta)$ to the new coordinates as follows:
\be\n{6.8a}
\psi_-=2\sqrt{2}e^{u_-(\theta)+w_-(\theta)-u_1-w_1}\,\tau^{1/2}\hhh\theta=\theta_{0}\,.  
\ee
In the coordinates  $(\tau,\theta_{0}=\theta)$ the metric \eq{6.1} takes the following form:
\be\n{6.9}
dS_{-}^2\approx -d\tau^2+\frac{1}{2\tau}e^{-4(u_1+w_1)}\,dT^2+2\tau e^{4(u_1+w_1)}\,d\Sigma^2_{-}\,,
\ee
where
\ba
d\Sigma^2_{-}&=&\frac{1}{4}\biggl(e^{-2(u_-(\theta)+w_-(\theta)+u_1+w_1)}d\theta^2\non\\
&+&2(1+\cos\theta)e^{2(u_-(\theta)-3u_1)}d\chi^2\non\\
&+&2(1-\cos\theta)e^{2(w_-(\theta)-3w_1)}d\phi^2\biggr)\,.\n{6.10}
\ea
The metric \eq{6.9} has the same Kasner exponents as those of \eq{6.7}.
 
The Kretschmann scalar \eq{6.3a} in the $(\tau,\theta_{0}=\theta)$ coordinates reads
\ba
&&\hspace{-0.7cm}{\cal K}_{-}\approx\frac{9}{2\tau^4}\left[1+\tilde{{\cal K}}_{-}^{(2)}\tau\right]\,,\n{6.11a}\\
&&\hspace{-0.7cm}\tilde{{\cal K}}_{-}^{(2)}=\frac{16}{3}e^{2(u_-(\theta)+w_-(\theta)-u_1-w_1)}\biggl[u_{-,\theta\theta}+w_{-,\theta\theta}\non\\
&&\hspace{-0.6cm}-\,4u_{-,\theta}^2-6u_{-,\theta}w_{-,\theta}-4w_{-,\theta}^2\non\\
&&\hspace{-0.6cm}-\,2\cot\theta(u_{-,\theta}+w_{-,\theta})+\left.\frac{u_{-,\theta}-w_{-,\theta}}{\sin\theta}+\frac{1}{2}\right].\n{6.11b}
\ea
We see that the expansion \eq{6.11a} coincides in the leading order with the expansion \eq{6.8}. Hence, in the presence of distortion, surfaces where the Kretschmann scalar has a constant value ${\cal K}_{-}={\cal K}_c$ are (in the leading order) surfaces of constant $\tau$. For $\tau\sim\tau_{\ind{Pl}}$ we can neglect the higher order terms in the expansion \eq{6.11a} and present the metric on the stretched singularity defined by ${\cal K}_{c}\sim\ell_{\ind{Pl}}^{-4}$ as follows:
\be\n{6.13}
dl_{-}^2\approx \left[\frac{{\cal K}_c}{72}\right]^{1/4}e^{-4(u_1+w_1)}\,dT^2+\left[\frac{72}{{\cal K}_c}\right]^{1/4}e^{4(u_1+w_1)}\,d\Sigma^2_{-}\, ,
\ee
where $d\Sigma^2_-$ is given by expression \eq{6.10}. According to the form of this metric, the stretched singularity of a distorted black hole has the same topology as the stretched singularity of a Schwarzschild-Tangherlini black hole.

\subsection{Geometry of the stretched singularity surface: duality transformation}

\begin{figure*}[htb]
\begin{center}
\ba
&&\hspace{-0.3cm}\includegraphics[width=5cm]{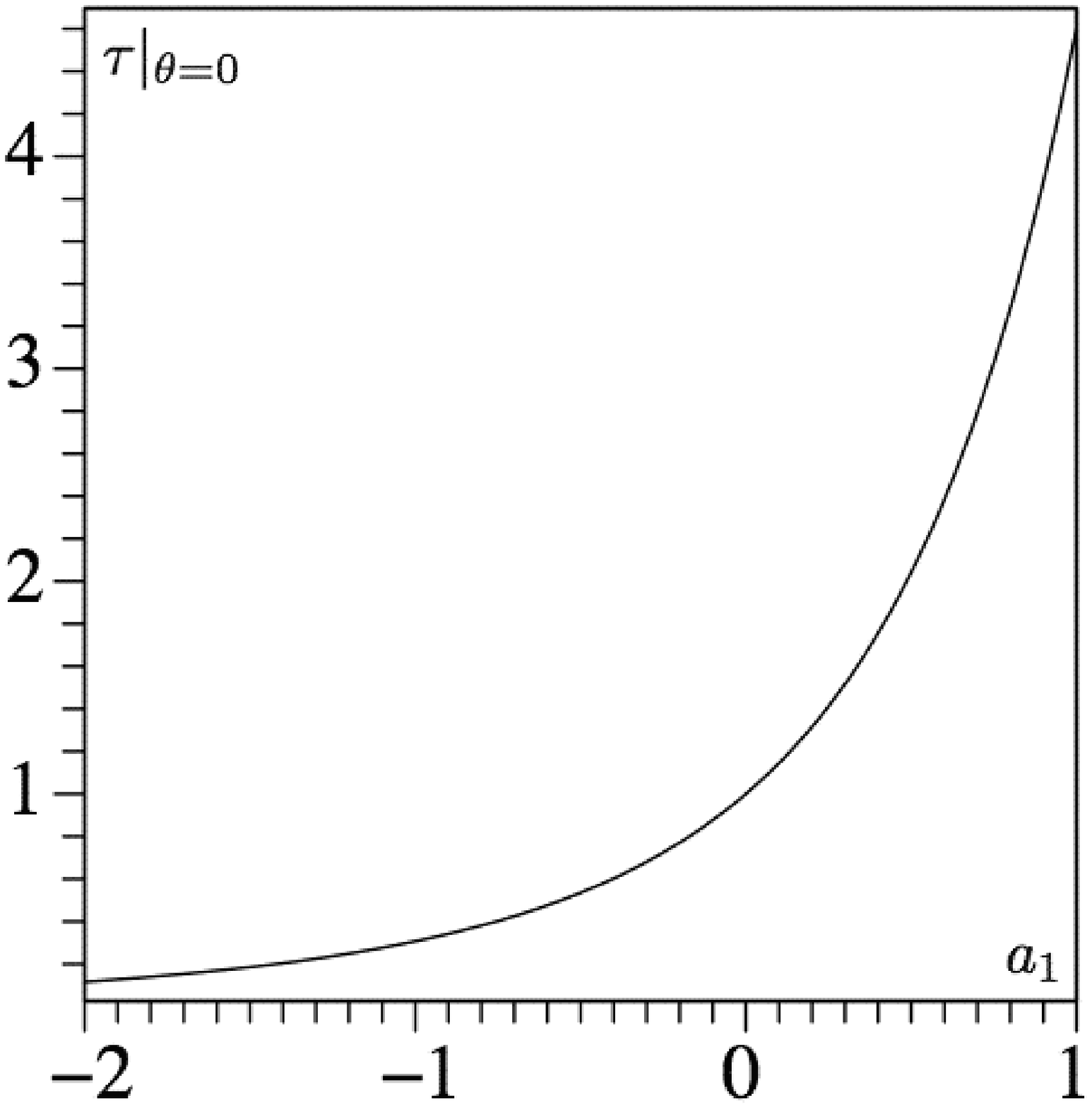}
\hspace{1cm}\includegraphics[width=5.15cm]{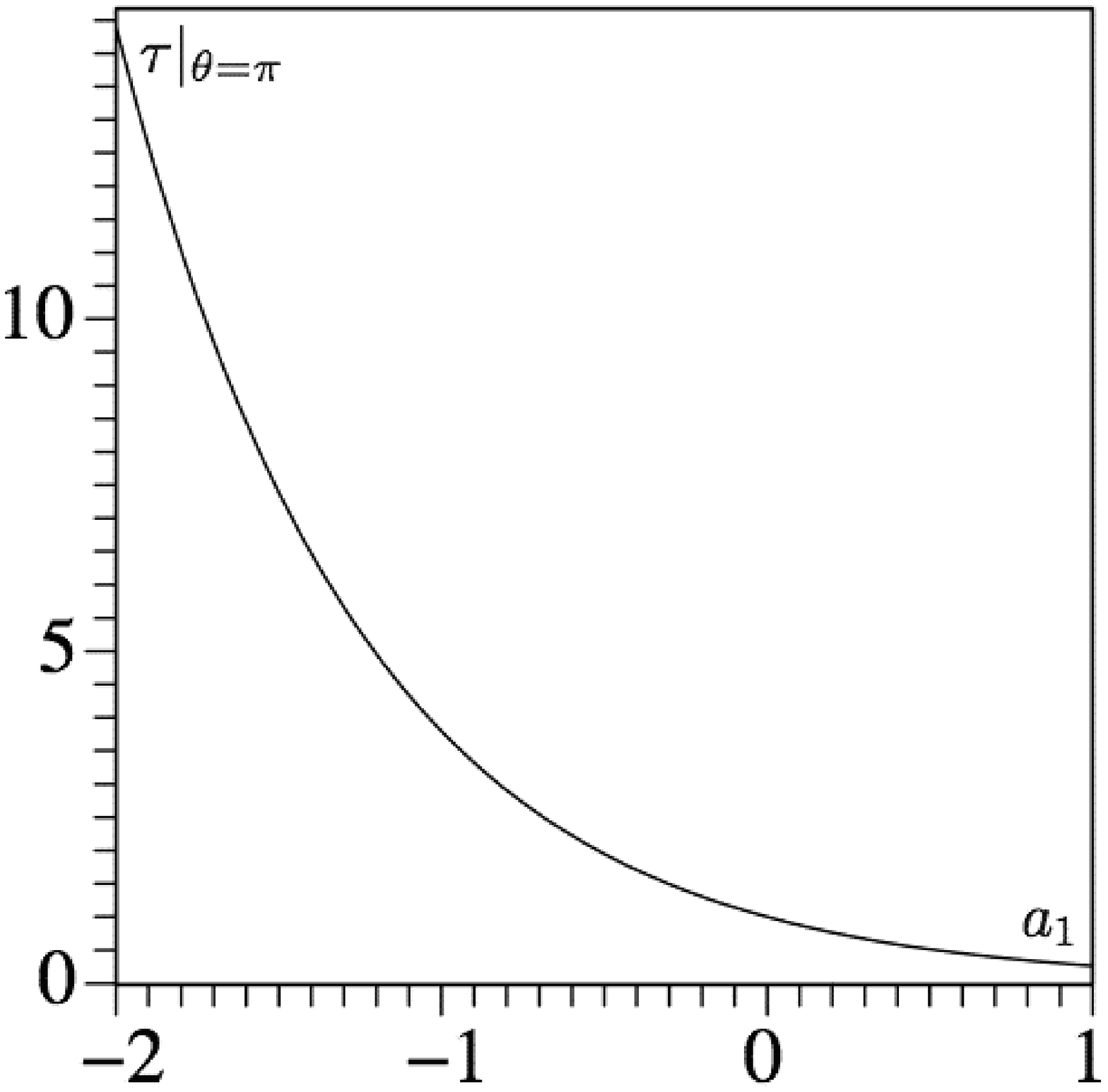}
\hspace{1cm}\includegraphics[width=5cm]{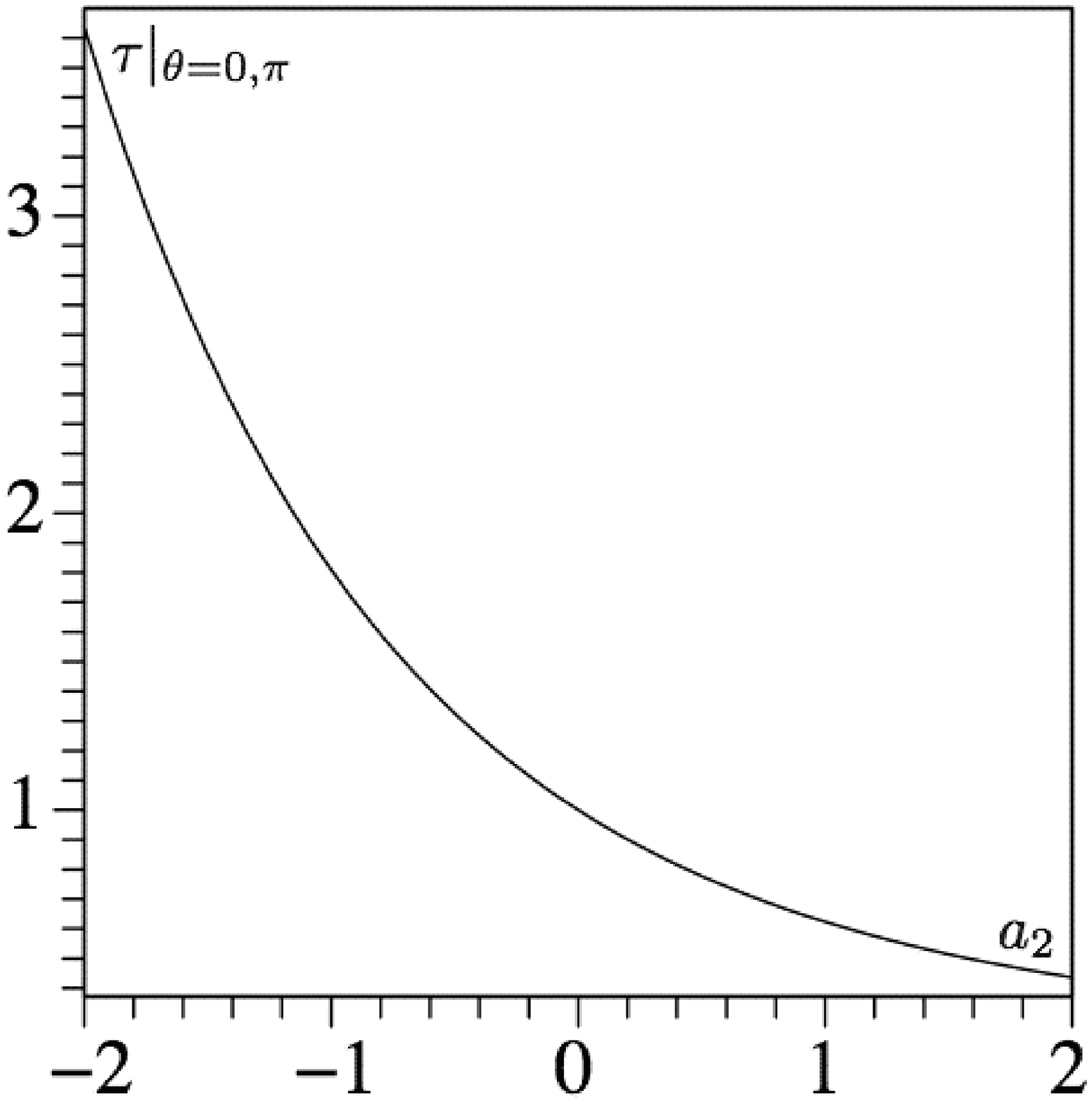}\non\\
&&\hspace{2.2cm}({\bf a})\hspace{5.6cm}({\bf b})\hspace{5.6cm}({\bf c})\non
\ea
\caption{The maximal proper time $\tau$ in units of $R_o$. ({\bf a}): The maximal proper time $\tau|_{\theta=0}$ for the dipole-monopole distortion. ({\bf b}): The maximal proper time $\tau|_{\theta=\pi}$ for the dipole-monopole distortion. ({\bf c}): The maximal proper time $\tau|_{\theta=0}=\tau|_{\theta=\pi}$ for the quadrupole-quadrupole distortion.}\label{F6}
\end{center} 
\end{figure*}
As we found in the previous subsection, the distortion fields do not change the topology of the stretched singularity of a Schwarzschild-Tangherlini black hole; however, they do change its geometry. To study the geometry of the stretched singularity, we consider the geometry of its 3-dimensional hypersurface defined by $T=const$. This surface is the Killing vector $\xi^\alpha_{(T)}$ orbit surface, i.e., it is invariant under $\mathbb{R}^{1}_{T}$ transformations. The metric on this surface is defined (up to the conformal factor) by $d\Sigma^2_-$ (see \eq{6.10}). We can calculate the intrinsic curvature of the stretched singularity surface and illustrate its shape by an isometric embedding of its 2-dimensional sections, as we did in Sec. VI for horizon surface of a distorted black hole. However, one can notice that the metric $d\Sigma^2_-$ can be obtained from the horizon surface metric $d\Sigma^2_+$ (see \eq{5.1}) by the following duality transformation:
\be
u_+\to -w_-\hhh u_1\to -w_1\hhh w_+\to -u_-\hhh w_1\to -u_1\,.\n{6.15} 
\ee 
The duality transformation corresponds to the exchange between the multipole moments (see Eqs. \eq{4.6a}, \eq{4.6b}, \eq{4.8a}, and \eq{4.8b})
\be\n{6.15a}
a_{2n+1}\longleftrightarrow b_{2n+1}\hhh a_{2n}\longleftrightarrow-b_{2n}\,.
\ee
The no-conical-singularity condition \eq{4.7} remains satisfied under the transformation \eq{6.15a}. The transformation \eq{6.15a} corresponds to the exchange between the ``axes" $\theta=0$ and $\theta=\pi$ and the reversal of the signs of the multipole moments (cf. \eq{tr1}):
\ba\n{tr2}
(\theta,\chi,\phi)&\to&(\pi-\theta,\phi,\chi)\,,\non\\
\\
\left[a_{n},b_{n}\right]&\to&[(-1)^{n+1}b_{n},(-1)^{n+1}a_{n}]\,.\non
\ea
Because the exchange between the ``axes" does not change the space-time of the distorted black hole, the transformation \eq{tr2} reduces to change of signs of the multipole moments. Thus, the stretched singularity intrinsic curvature invariants can be derived from those of the horizon surface illustrated in Fig.~\ref{F3} by exchanging Lines 1, 2, 3, and 4, with Lines 2, 1, 4, and 3, and changing $\theta$ to $\pi-\theta$ in each of Figs.~\ref{F3} ({\bf a}) and ({\bf b}). An isometric embedding of the stretched singularity sections can be derived from those of the horizon surface illustrated in Fig.~\ref{F4} by exchanging Lines 1, 2, 3, and 4 in Fig.~\ref{F4} ({\bf a}) with Lines 2, 1, 4, and 3 in Fig.~\ref{F4} ({\bf b}).

According to the duality transformation, given a 5-dimensional distorted black hole, one can find another one whose horizon surface geometry is the same as the geometry of the stretched singularity of the former one.  

\section{Proper time of free fall from the horizon to the singularity}

So far we have studied the effect of distortion on the black hole horizon and singularity. However, the distortion fields affect the entire interior region of the black hole. To illustrate this, we consider how the distortion fields change the proper time of free fall of a test particle moving from the horizon to the singularity. 

Namely, we study how the proper time depends on the multipole moments of the distortion fields. In our study we consider adiabatic distortion, so that the area $A_{+}$ of the distorted black hole horizon surface remains constant, which is equal to the horizon surface area of an undistorted Schwarzschild-Tangherlini black hole. We define the proper time in units of the radius $R_{o}$ corresponding to the area $A_{+}$ (see \eq{dar}),
\be
R_{o}=\left(\frac{A_{+}}{2\pi^{2}}\right)^{1/3}=r_{o}e^{-(u_{1}+w_{1}+3u_{0}+3w_{0})/6}\,.
\ee
To make our analysis simpler, we consider a test particle moving along a timelike geodesic defined by $(t,\chi,\phi)=const$ and with $L_0=0$  (see Appendix D). Such a ``radial'' motion corresponds to zero angular momenta and energy of the particle. One can show that the proper time is maximal for such a motion. In addition, we consider free fall from the horizon to the singularity along each of the symmetry ``axes" $\theta=0$ and $\theta=\pi$. Using the metric \eq{4.2} we derive
\ba
&&\hspace{-0.7cm}\tau|_{\theta=0}=\frac{r_o}{2\sqrt{2}R_{o}}\int_{0}^{\pi}(1+\cos\psi)^{1/2}e^{-u_+(\psi)-u_0}d\psi\,,\n{7.1a}\\
&&\hspace{-0.7cm}\tau|_{\theta=\pi}=\frac{r_o}{2\sqrt{2}R_{o}}\int_{0}^{\pi}(1+\cos\psi)^{1/2}e^{-w_-(\psi)-w_0}d\psi\,.\n{7.1b}
\ea
For a Schwarzschild-Tangherlini black hole the maximal proper time of free fall along a radial timelike geodesic is equal to $1$ in units of $R_{o}=r_{o}$. 

Let us now calculate the maximal proper time for the dipole-monopole distortion \eq{3.11a}. In this case the integrals \eq{7.1a} and \eq{7.1b} can be evaluated exactly,
\ba
\tau|_{\theta=0}&=&\frac{\sqrt{\pi}}{2\sqrt{-2a_1}}e^{2a_1/3}\text{erf}(\sqrt{-2a_1})\,,\n{7.2a}\\
\tau|_{\theta=\pi}&=&e^{-4a_1/3}\,.\n{7.2b}
\ea
Here $\text{erf}(x)$ is the error function (see, e.g., \cite{AS}, p. 297). The maximal proper time for the quadrupole-quadrupole distortion \eq{3.12a} is the same for free fall along both the ``axes",
\be\n{7.3}
\tau|_{\theta=0}=\tau|_{\theta=\pi}=\int_0^1e^{-4a_2(x^2-x^4)}dx\,.
\ee
The maximal proper time calculated for the black hole distorted by the dipole-monopole and quadrupole-quadrupole distortions is shown in Figs.~\ref{F6}({\bf a}), ({\bf b}), and ({\bf c}). According to these figures, for some values of the multipole moments the maximal proper time is less, equal, or greater than that of a Schwarzschild-Tangherlini black hole of the same horizon area. One can see that due to the external distortion, the singularity of a Schwarzschild-Tangherlini black hole can come close to its horizon.

\section{Summary of Results and Discussion}

In this paper we studied a distorted, 5-dimensional vacuum black hole as a 5-dimensional Schwarzschild-Tangherlini black hole distorted by a static, neutral external distribution of matter. We constructed the corresponding metric which represents such a {\em local black hole}. In other words, the distortion sources are not included into the metric but are put at infinity. The metric is presented in the 5-dimensional Weyl form which admits the $\mathbb{R}^{1}\times U(1)\times U(1)$ isometry group. This metric is a 5-dimensional generalization of the 4-dimensional Weyl form representing the corresponding distorted vacuum black hole studied before (see, e.g., \cite{Geroch,Chandrabook,FN,Dor,MySz,Werner,FS}). 

As a result of our study, we found that distortion fields affect the black hole horizon and singularity. The 5-dimensional distorted black hole has the following properties, which are common with the 4-dimensional one: There is a certain duality transformation between the black hole horizon and the {\em stretched singularity} surfaces. This transformation implies that distortion of the horizon surface uniquely defines distortion of the stretched singularity surface. Given a 5-dimensional black hole, one can ``observe'' its distorted stretched singularity surface by observing the horizon surface of the dual distorted black hole. The topology of the stretched singularity is the same as that of a Schwarzschild-Tangherlini black hole. Moreover, the Kasner exponents of the space-time region near the singularities of the black holes are the same as well. One may assume that these properties are the inherent properties of the 4- and 5-dimensional Weyl forms, representing such distorted black holes. Whether all or some of these properties will remain if one changes the $U(1)\times U(1)$ symmetry (for example to $SO(3)$) remains an open question. Thus, we cannot say if a 5-dimensional {\em compactified} black hole has similar properties. However, a 4-dimensional  compactified black hole indeed has properties similar to those of a 4-dimensional distorted black hole \cite{FF,FS}.

The analysis of the maximal proper time of free fall from the distorted black hole horizon to its singularity along the symmetry ``axes" shows that the proper time can be less, equal to, or greater than that of a Schwarzschild-Tangherlini black hole of the same horizon area. As a result of external distortion, the black hole stretched singularity can approach the horizon. In particular, the black hole stretched singularity approach its horizon. This scenario may suggest that the singularity of a 5-dimensional compactified black hole can approach its horizon during an infinitely slow merger transition between the black hole and the corresponding black string\footnote[5]{When the ``north'' and the ``south'' poles of a 4-dimensional compactified black hole come close to each other during an infinitely slow merger transition, its stretched singularity becomes naked at the vicinity of the poles \cite{FS}.}. If so, one cannot rely on a classical description of the transition. 

In our paper we derived a relation between the Kretschmann scalar calculated on the horizon of a 5-dimensional static, asymmetric, distorted vacuum black hole and the trace of the square of the Ricci tensor of the horizon surface. This relation is a generalization of a similar relation between the Kretschmann scalar calculated on the horizon of a 4-dimensional static, asymmetric, distorted vacuum black hole and the square of the Gaussian curvature of its horizon surface (see \cite{FS} and \cite{FSAn}).

Our construction and analysis of a 5-dimensional distorted black hole is based on the 5-dimensional Weyl form (see Sec. II), which is adopted for the construction of 5-dimensional black objects distorted by external gravitational fields. Using this Weyl form one can study other 5-dimensional black objects, e.g., distorted black strings and black rings. One can consider distorted higher($>$5)-dimensional black objects as well, by using the corresponding Weyl form.

\appendix

\section{Space-time invariants}

In this appendix we derive a relation between the Kretschmann scalar ${\cal K}$ calculated on the horizon of a 5-dimensional static, asymmetric, distorted vacuum black hole and the trace of the square of the Ricci tensor, ${\cal R}_{AB}{\cal R}^{AB}$, of the horizon surface. The corresponding space-time admits the Killing vector $\xi^{\alpha}=\delta^{\alpha}_{\,\,\, 0}$, ($x^0:=t$), which is timelike in the domain of interest, $\xi^{\alpha}\xi_{\alpha}=g_{00}:=-k^2<0$, and hypersurface orthogonal. The space-time metric $g_{\alpha\beta}$, $(\alpha,\beta,...=0,\ldots,4)$ can be presented in the form
\ba\n{A1}
&&\hspace{-1cm}ds^2=g_{\alpha\beta}dx^{\alpha}dx^\beta=-k^2dt^2+\gamma_{ab}(x^c)dx^adx^b\,,
\ea
where $\gamma_{ab}$, $a,b,c,...=1,\ldots,4$, is the metric on a 4-dimensional hypersurface $t=const$. The black hole horizon defined by $k=0$ is a non-degenerate Killing horizon. One can show that the vacuum Einstein equations $^{(5)}\hspace{-0.09cm}R_{\alpha\beta}=0$ for the metric \eq{A1} reduce to\footnote[6]{One can derive the Einstein equations \eq{A2} and \eq{A3} starting from the 5-dimensional vacuum Einstein equations and using Eqs. \eq{A6} and \eq{A8} adopted to the metric \eq{A1} (for details see, e.g., \cite{Israel}).}
\ba
&&\hspace{-0.2cm}R_{ab}-k^{-1}\nabla_a\nabla_bk=0\,,\n{A2}\\
&&\hspace{0.5cm}\nabla_a\nabla^ak=0\,,\n{A3}
\ea
where $R_{ab}$ and $\nabla_a$ are the Ricci tensor and the covariant derivative defined with respect to the 4-dimensional metric $\gamma_{ab}$. Equation \eq{A3} implies that $k$ is a harmonic function. Thus, $k$ can be considered in each 4-dimensional hypersurface $t=const$. As a result, the metric \eq{A1} can be written in the following form:
\ba\n{A4}
ds^2&=&-k^2dt^2+\kappa^{-2}(k,x^C)dk^2+h_{AB}(k,x^C)dx^Adx^B\,,\non\\
\ea
where $h_{AB}$, $A,B,C,...=1,2,3$, is the metric on an orientable 3-dimensional hypersurface $\Sigma_k$. One can show that
\be\n{A5}
\kappa^2(k,x^C)=-\frac{1}{2}(\nabla^{\alpha}\xi^{\beta})(\nabla_{\alpha} \xi_{\beta})\,,
\ee
where $\nabla_{\alpha}$ is the covariant derivative defined with respect to the metric \eq{A1}. Thus, $\kappa(k=0,x^C)$ coincides with the surface gravity of a 5-dimensional vacuum black hole.
 
To present geometric quantities of the 5-dimensional space-time \eq{A4} in terms of these corresponding to $\Sigma_k$, we use the following relations:
\ba
R_{ABCD}&=&{\cal R}_{ABCD}+{\cal S}_{AD}{\cal S}_{BC}-{\cal S}_{AC}{\cal S}_{BD}\,,\n{A6}\\
R_{kABC}&=&\kappa^{-1}({\cal S}_{AB;C}-{\cal S}_{AC;B})\,,\n{A7}\\
R_{AkkB}&=&\kappa^{-1}(h_{AC}{\cal S}_{B\,\,\,,k}^{\,\,\,\,C}+(\kappa^{-1})_{;AB}+\kappa^{-1}{\cal S}_{AC}{\cal S}_{B}^{\,\,\,\,C})\,,\non\\
\n{A8}
\ea
where the first two equations are due to Gauss and Codazzi (see, e.g., \cite{Israel} and \cite{Eis}). Here ${\cal R}_{ABCD}$ is the intrinsic curvature, and
\be\n{A9}
{\cal S}_{AB}=\frac{\kappa}{2} h_{AB,k}
\ee
is the extrinsic curvature of a hypersurface $\Sigma_k$. The semicolon $;$ stands for the covariant derivative defined with respect to the metric $h_{AB}$. 

Using expressions \eq{A4} and \eq{A9}, we derive
\ba\n{A10}
&&\hspace{-0.3cm}\nabla_k\nabla_kk=\kappa^{-1}\kappa_{,k}\hhh \nabla_A\nabla_kk=\nabla_k\nabla_Ak=\kappa^{-1}\kappa_{,A}\,,\non\\ 
&&\hspace{-0.3cm}\nabla_A\nabla_Bk=\kappa{\cal S}_{AB}\hhh \nabla_a\nabla^ak=\kappa (\kappa_{,k}+{\cal S})\hhh {\cal S}\equiv {\cal S}_{A}^{\,\,\,\,A}\,.\non\\
\ea
Applying expressions \eq{A6}-\eq{A10} to the Einstein equations \eq{A2} and \eq{A3}, we derive the following set of equations:
\ba
&&\hspace{-0.7cm}\kappa_{,k}+{\cal S}=0\,,\n{A11}\\
&&\hspace{-0.7cm}{\cal R}_{A}^{\,\,\,\,B}=\kappa {\cal S}_{A\,\,\,,k}^{\,\,\,\,B}+\kappa(\kappa^{-1})_{;A}^{\,\,\,\,\,;B}+{\cal S}{\cal S}_{A}^{\,\,\,\,B}+k^{-1}\kappa {\cal S}_{A}^{\,\,\,\,B}\,,\n{A12}\\
&&\hspace{-0.7cm}\kappa_{,A}+k({\cal S}_{,A}-{\cal S}_{A\,\,\,;B}^{\,\,\,\,B})=0\,,\n{A13}\\
&&\hspace{-0.7cm}{\cal R}:=h^{AB}{\cal R}_{AB}={\cal S}^2-S_{AB}{\cal S}^{AB}+2k^{-1}\kappa {\cal S}\,,\n{A14}\\
&&\hspace{-0.7cm}{\cal S}_{,k}+(\kappa^{-1})_{;A}^{\,\,\,\,\,;A}+\kappa^{-1}{\cal S}_{AB}{\cal S}^{AB}+k^{-1}\kappa_{,k}=0\,.\n{A15}
\ea
Equations \eq{A9}, \eq{A11}, and \eq{A12} define a complete system for determining $\kappa$, $h_{AB}$, and ${\cal S}_{AB}$ as functions of $k$. The constraint equations \eq{A13} and \eq{A14} together with Eq. \eq{A15} are satisfied for any value of $k$.

For the static space-time \eq{A1}, the Riemann tensor components are given by\footnote[7]{Expressions \eq{A16} can be derived by changing notations in expressions \eq{A6}-\eq{A8} as follows: $k\to t$, $\kappa\to ik^{-1}$, and taking into account that the extrinsic curvature of a 4-dimensional hypersurface $t=const$ vanishes (see Eq. \eq{A9}).}
\ba\n{A16}
^{(5)}\hspace{-0.09cm}R_{attb}&=&-k\nabla_a\nabla_bk \hhh ^{(5)}\hspace{-0.09cm}R_{tabc}=0\hhh ^{(5)}\hspace{-0.09cm}R_{abcd}=R_{abcd}\,.\non\\
\ea
Thus, we arrive to the following expression for the Kretschmann scalar of the space-time \eq{A4}:
\ba\n{A17}
{\cal K}&\equiv&^{(5)}\hspace{-0.09cm}R_{\alpha\beta\gamma\delta}\hspace{0.05cm}^{(5)}\hspace{-0.09cm}R^{\alpha\beta\gamma\delta}=4k^{-2}(\nabla_a\nabla_bk)(\nabla^a\nabla^bk)\non\\
&+&4R_{AkkB}R^{AkkB}+4R_{kABC}R^{kABC}\non\\
&+&R_{ABCD}R^{ABCD}\,.
\ea
Let us present this expression in terms of 3-dimensional geometric quantities defined on $\Sigma_k$. Using Eq. \eq{A6} we derive
\ba\n{A18}
R_{ABCD}R^{ABCD}&=&{\cal R}_{ABCD}{\cal R}^{ABCD}\non\\
&+&2{\cal R}_{ABCD}({\cal S}^{AD}{\cal S}^{BC}-{\cal S}^{AC}{\cal S}^{BD})\non\\
&+&2({\cal S}_{AB}{\cal S}^{AB})^2-2{\cal S}_{AC}{\cal S}^{BC}{\cal S}^{AD}{\cal S}_{BD}\,.\non\\
\ea
The 3-dimensional Riemann tensor components ${\cal R}_{ABCD}$ corresponding to $h_{AB}$ can be presented as follows (see, e.g., \cite{MTW}, p. 550):
\ba\n{A19}
{\cal R}_{ABCD}&=&h_{AC}{\cal R}_{BD}+h_{BD}{\cal R}_{AC}-h_{AD}{\cal R}_{BC}\non\\
&-&h_{BC}{\cal R}_{AD}-\frac{1}{2}{\cal R}(h_{AC}h_{BD}-h_{AD}h_{BC})\,,\non\\
\ea
where the Ricci scalar ${\cal R}$ and the trace of the square of the Ricci tensor ${\cal R}_{AB}{\cal R}^{AB}$ are defined on $\Sigma_k$. Expression \eq{A19} implies
\ba\n{A20}
{\cal R}_{ABCD}{\cal R}^{ABCD}&=&4{\cal R}_{AB}{\cal R}^{AB}-{\cal R}^2\,.
\ea
Using expressions \eq{A7}, \eq{A8}, \eq{A10}, \eq{A17}, \eq{A18}, \eq{A19}, and \eq{A20} we derive
\ba\n{A21}
{\cal K}&=&4k^{-2}\kappa^2\left(\kappa_{,k}^2+2\kappa^{-2}\kappa_{,A}\kappa^{,A}+2{\cal S}_{AB}{\cal S}^{AB}\right)\non\\
&-&8k^{-1}\kappa {\cal S}^{AB}({\cal R}_{AB}-{\cal S}{\cal S}_{AB}+{\cal S}_{AC}{\cal S}_{B}^{\,\,\,\,C})\non\\
&+&8{\cal R}_{AB}{\cal R}^{AB}-{\cal R}^2+2{\cal S}^2({\cal R}+2{\cal S}_{AB}{\cal S}^{AB})\non\\
&-&2{\cal S}_{AB}{\cal S}^{AB}({\cal R}-{\cal S}_{CD}{\cal S}^{CD})\non\\
&-&8{\cal S}{\cal S}^{AB}(2{\cal R}_{AB}+{\cal S}_{AC}{\cal S}_{B}^{\,\,\,\,C})\non\\
&+&2{\cal S}^{AC}{\cal S}^{B}_{\,\,\,\,C}(8{\cal R}_{AB}+{\cal S}_{AD}{\cal S}_{B}^{\,\,\,\,D})\non\\
&+&8{\cal S}^{AB;C}({\cal S}_{AB;C}-{\cal S}_{AC;B})\,.
\ea
Thus, one can see that the black hole horizon $k=0$ is regular if $\kappa_{,A}=0$ and ${\cal S}_{AB}=0$ on the horizon, i.e., the surface gravity is constant on the horizon, and the horizon surface, defined by $k=0$ and $t=const$, is a totally geodesic surface which is regular, i.e., ${\cal R}_{AB}{\cal R}^{AB}$ and ${\cal R}$ are finite on the surface.

To derive a relation between the Kretschmann scalar calculated on the horizon and the 3-dimensional geometric quantities defined on the horizon surface, we use the following series expansions:
\ba\n{A22}
{\cal A}&=&\sum_{n\geqslant0}{\cal A}^{(2n)}k^{2n}\hhh {\cal B}=\sum_{n\geqslant0}{\cal B}^{(2n+1)}k^{2n+1}\,,
\ea
where ${\cal A}:=\{h_{AB},\kappa,{\cal R}_{AB},{\cal R}\}$ and ${\cal B}:=\{{\cal S}_{AB},{\cal S}\}$. Here the first term ${\cal A}^{(0)}$ corresponds to the value of ${\cal A}$ calculated on the horizon. To calculate ${\cal K}$ on the horizon it is enough to consider the first order expansion only, i.e., $n=0,1$.  Substituting expansions \eq{A22} into equations \eq{A9}, \eq{A11}-\eq{A15}, we derive
\ba\n{A23}
&&\hspace{-0.8cm}(\kappa^{(0)})_{,A}=0\hhh \kappa^{(2)}=-\frac{{\cal R}^{(0)}}{4\kappa^{(0)}}\hhh {\cal S}_{AB}^{(1)}=\frac{{\cal R}_{AB}^{(0)}}{2\kappa^{(0)}}\,,\non\\
&&\hspace{-0.8cm}{\cal S}^{(1)}=h^{AB(0)}{\cal S}_{AB}^{(1)}=\frac{{\cal R}^{(0)}}{2\kappa^{(0)}}\hhh h_{AB}^{(2)}=\frac{{\cal R}_{AB}^{(0)}}{2(\kappa^{(0)})^{2}}\,.
\ea  
Substituting the corresponding expansions \eq{A22} for $n=0,1$ with the coefficients \eq{A23} into Eq. \eq{A21}, we derive the following relation between the Kretschmann scalar ${\cal K}$ calculated on the horizon of a 5-dimensional static, asymmetric, distorted vacuum black hole and the trace of the square of the Ricci tensor ${\cal R}_{AB}{\cal R}^{AB}$ of the horizon surface:
\ba\n{A24}
{\cal K}_+=6({\cal R}_{AB}{\cal R}^{AB})_+\,.
\ea
It is interesting to note that the same relation holds for 4-dimensional static space-times. Namely, if we consider ${\cal R}_{AB}$ as the Ricci tensor of the 2-dimensional horizon surface of a 4-dimensional static asymmetric black hole, then the relation becomes ${\cal K}_+=3{\cal R}_+^2$ (see, e.g., \cite{FS} and \cite{FSAn}).

\section{Gaussian curvatures}

The Gaussian curvatures \eq{K1}--\eq{K3} corresponding to the 3-dimensional horizon surface defined by the metric \eq{5.1} are the following:
\ba
K_{+\phi}&=&{\cal N}\biggl(1+4w_{+,\theta\theta}-8w_{+,\theta}^2-4u_{+,\theta}w_{+,\theta}\non\\
&-&\frac{2\sin\theta}{1+\cos\theta}(u_{+,\theta}+3w_{+,\theta})\biggr)\,,\n{KP1}\\
K_{+\chi}&=&{\cal N}\biggl(1+4u_{+,\theta\theta}-8u_{+,\theta}^2-4u_{+,\theta}w_{+,\theta}\non\\
&+&\frac{2\sin\theta}{1-\cos\theta}(w_{+,\theta}+3u_{+,\theta})\biggr)\,,\n{KX1}\\
K_{+\theta}&=&{\cal N}\biggl(1-4u_{+,\theta}w_{+,\theta}-\frac{2}{\sin\theta}(u_{+,\theta}-w_{+,\theta})\non\\
&+&2\cot\theta(u_{+,\theta}+w_{+,\theta})\biggr)\,,\n{KT1}
\ea
where ${\cal N}=e^{-2(u_+(\theta)+w_+(\theta)+u_1+w_1)}$. 

\section{Distortion fields $\hu$, $\hw$, and $\hv$ near the black hole horizon and singularity}

To study the behavior of the distortion fields $\hu$, $\hw$, and $\hv$ near the distorted black hole horizon and singularity, it is convenient to use the $\psi$ coordinate (see \eq{4.1}). We can expand the distortion fields given by the exact solutions \eq{3.3a}--\eq{3.4c} of the Einstein equations near the black hole horizon and singularity. However, to derive a simple form of such expansions, it is easy to construct an approximate solutions to the Einstein equations \eq{2.15}--\eq{2.16b}. Using Eq. \eq{4.1}, we present Eq. \eq{2.15} in the following form:
\be\n{C1}
D_{\psi}\hx(\psi,\theta)=D_\theta\hx(\psi,\theta)\hhh \hx:=(\hu,\hw)\,,
\ee
where 
\be\n{C2}
D_\sigma:=\partial^2_\sigma+\cot\sigma\partial_\sigma\hhh \sigma:=(\psi,\theta)\,.
\ee
The black hole horizon and singularity correspond to $\psi=0$ and $\psi=\pi$, respectively. To consider both the cases simultaneously we denote $\psi_{+}:=\psi-0=\psi$ and $\psi_-:=\pi-\psi$. According to Eq. \eq{3.3a}, the function $\hx$ is an even function of $\psi_\pm$. Thus, near the horizon and the singularity it has the following expansion:
\be\n{C3}
\hx(\psi,\theta)=\sum_{k=0}^\infty X_{\pm}^{(2k)}(\theta)\,\psi_{\pm}^{2k}\,.
\ee
Using the series expansion for $\cot \psi_{\pm}$ (see, e.g., \cite{AS}, p. 75),
\ba\n{C4}
\cot\psi_{\pm}&=&\psi_{\pm}^{-1}\left[1-\sum_{m=1}^{\infty} C_{2m}\psi_{\pm}^{2m}\right]\,,\\
C_{2m}&=&\frac{(-1)^{m-1} 2^{2m} B_{2m}}{(2m)!}\hhh |\psi_\pm|<\pi\,,
\ea
where $B_{2m}$ are the Bernoulli numbers
\be\n{C5}
B_2=\frac{1}{6}\hhh B_4=-\frac{1}{30}\hhh B_6=\frac{1}{42}\, \ldots \,,
\ee
we derive
\be\n{C6}
D_{\psi_{\pm}}\psi_{\pm}^{2k}=4k^2\psi_{\pm}^{2(k-1)}-
2k\sum_{m=1}^{\infty}C_{2m}\psi_{\pm}^{2(k+m-1)}\,. 
\ee
Substituting expansion \eq{C3} into Eq. \eq{C1} and using Eq. \eq{C6}, we derive the following recurrence relations for $X_{\pm}^{(2k)}(\theta)$:
\ba
X_{\pm}^{(0)}&=&x_\pm(\theta)+x_0\,,\non\\
X_{\pm}^{(2)}&=&\frac{1}{4}(x_{\pm,\theta\theta}+\cot\theta x_{\pm,\theta})\,,\n{C7a}\\
&\vdots&\non\\
X_{\pm}^{(2k+2)}&=&\frac{1}{4(k+1)^2}\left[D_{\theta}X_{\pm}^{(2k)}\right.\non\\
&+&\left.2\sum_{m=1}^{k} (k-m+1) C_{2m}X_{\pm}^{(2(k-m+1))}\right]\,,\non\\
k&=&0,1,2,\ldots\,.\n{C7b} 
\ea 
Here $x_\pm:=(u_\pm,w_\pm)$ and $x_0:=(u_0,w_0)$ (see Eqs. \eq{4.6a}, \eq{4.6b}, \eq{4.8a}, and \eq{4.8b}).

The asymptotic expansion of the distortion field $\hv$, which is an even function of $\psi_\pm$, near the horizon and the singularity, can be written in the form
\be\n{C8}
\hv(\psi,\theta)=\sum_{k=0}^{\infty} V_{\pm}^{(2k)}(\theta)\,\psi_{\pm}^{2k}\,.
\ee
Substituting this expansion together with expansion \eq{C3} of the distortion fields $\hu$ and $\hw$ into equation \eq{2.16a} (with $\eta$ replaced by $\psi$, according to \eq{4.1}) we can determine the functions $V_{\pm}^{(2k)}(\theta)$. The first two of these functions are the following:
\ba
V_{+}^{(0)}&=&-\frac{3}{2}(u_0+w_0)-\frac{1}{2}(u_1+w_1)\,,\non\\
V_{+}^{(2)}&=&\frac{1}{4}(2u_{+,\theta}^2+u_{+,\theta}w_{,+\theta}+w_{+,\theta}^2)-\frac{u_{+,\theta}-w_{+,\theta}}{4\sin\theta}\,\non\\
&-&\frac{3}{4}\cot\theta(u_{+,\theta}+w_{+,\theta})\,,\n{C9a}\\
&\vdots&\non\\
V_{-}^{(0)}&=&\frac{1}{2}[u_1+w_1-3(u_0+w_0)]-3(u_-(\theta)+w_-(\theta))\,,\non\\
V_{-}^{(2)}&=&\frac{1}{4}(2u_{-,\theta}^2+u_{-,\theta}w_{-,\theta}+w_{-,\theta}^2)-\frac{u_{-,\theta}-w_{-,\theta}}{4\sin\theta}\,\non\\
&-&\frac{3}{4}(u_{-,\theta\theta}+w_{-,\theta\theta})\,,\n{C9b}\\
&\vdots&\non
\ea

\section{Geodesics near the singularity}

For a free particle moving in a 5-dimensional distorted black hole interior there exist three integrals of motion related to the Killing vectors \eq{2.4}, the energy 
\be\n{D1a}
E:=-p_T=-\xi^\alpha_{(T)}p_{\alpha}\,,
\ee
and the angular momenta
\be\n{D1b}
L_{\chi}:=p_{\chi}=\xi^\alpha_{(\chi)}p_{\alpha}\hhh L_{\phi}:=p_{\phi}=\xi^\alpha_{(\phi)}p_{\alpha}\,.
\ee
which correspond to the ``axes" $\theta=\pi$ and $\theta=0$, respectively. The other five constants of motion that characterize geodesic motion in the black hole interior are $L_0$, the limiting value of
$L=[r_0\sin(\psi_-/2)]^2\dot{\theta}$ at the singularity $\psi_- =0$ (with $\psi_{-} =\pi-\psi$), and $\theta_0$, $t_0$, $\chi_0$, and $\phi_0$, the limiting values of $\theta$, $t$, $\chi$, and $\phi$, respectively, at the singularity.  For the Schwarzschild-Tangherlini black hole metric (21), $L = r^2\dot{\theta}$ is a constant of motion, but for a distorted black hole it is not.  However, it does have a finite limiting value $L_0$ at the singularity that may be taken to be a characteristic value for the entire geodesic and hence a constant of motion.

Consider an initial point with coordinates $(\psi_{-i}, \theta_i, t_i,\chi_i, \phi_i)$ near the singularity of the distorted black hole ($\psi_{-i} \ll 1$).  The proper time $\tau$ to fall from this point to the singularity depends on the location of the point and also on the geodesic constants of motion $E$, $L_\chi$, $L_\phi$, and $L_0$.  One can show that the maximal proper time from the point to the singularity corresponds to $E = L_\chi = L_\phi = L_0 = 0$. We shall call the corresponding geodesic ``radial''. For the ``radial'' geodesic, $(t,\chi,\phi)=const$ along the geodesic, so $t_0 = t_i$, $\chi_0 = \chi_i$, and $\phi_0 = \phi_i$.  In the Schwarzschild-Tangherlini black hole, $\theta$ would also be constant for a radial geodesic (which has $L=0$ all along it), so there $\theta_0 = \theta_i$, but for a distorted black hole neither $L$ nor $\theta$ is constant, so $\theta_0 \neq\theta_i$, though $\theta_0$ is uniquely determined by the initial point $(\psi_{-i}, \theta_i, t_i, \chi_i, \phi_i)$ and is actually a function only of $\psi_{-i}$ and $\theta_i$ for a fixed distorted black hole metric.  This ``radial'' geodesic is a geodesic of the 2-dimensional metric
\be\n{D2}
d\gamma^2=B_-(d\theta^2-d\psi_-^2)\,,
\ee
obtained by the dimensional reduction $(T,\chi,\phi)=const$ of the metric \eq{6.1}.

The Christoffel symbols for the metric \eq{D2} are
\ba
{}&&\Gamma^{\psi_-}_{\,\,\,\,\,\psi_-\psi_-}=\Gamma^{\theta}_{\,\,\,\,\theta\psi_-}
=\Gamma^{\psi_-}_{\,\,\,\,\,\theta\theta}=\frac{B_{-,\psi_-}}{2B_-}\,,\non\\
{}&&\Gamma^{\theta}_{\,\,\,\,\psi_-\psi_-}=\Gamma^{\theta}_{\,\,\,\,\theta\theta}
=\Gamma^{\psi_-}_{\,\,\,\,\,\theta\psi_-}=\frac{B_{-,\theta}}{2B_-}\,.\n{D4}
\ea
Thus, the geodesic equation
\be\n{D5}
\ddot{x}^{\alpha}+\Gamma^{\alpha}_{\,\,\,\beta\gamma}\,\dot{x}^{\beta}\dot{x}^{\gamma}=0
\ee
for the metric \eq{D2} takes the following form:
\ba
&&\hspace{-0.5cm}2B_{-}\ddot{\psi}_-+B_{-,\psi_-}(\dot{\psi}_-^2+\dot{\theta}^2)+2B_{-,\theta}\dot{\psi}_-\dot{\theta}=0\,,\n{D6a}\\
&&\hspace{-0.5cm}2B_{-}\ddot{\theta}+B_{-,\theta}(\dot{\psi}_-^2+\dot{\theta}^2)+2B_{-,\psi_-}\dot{\psi}_-\dot{\theta}=0\,.\n{D6b}
\ea
Here the over dot denotes the derivative with respect to the proper time $\tau$. These equations obey the constraint
\be
B_-(\dot{\psi}_-^2-\dot{\theta}^2)=1\,,\n{D7}
\ee
that is, the normalization condition $u_{\alpha}u^{\alpha}=-1$ for the $5$-velocity $u^{\alpha}$.

Expansion \eq{6.2} for the metric function $B_{-}$ near the singularity in the leading order in $\psi_-$ is
\be\n{D8}
B_-\approx \frac{\psi_-^2}{16}e^{-4(u_-(\theta)+w_-(\theta)-u_1-w_1)}\,.
\ee
Substituting this expression into the geodesic equations \eq{D6a}, \eq{D6b}, and the constraint \eq{D7}, we derive
\ba
&&\hspace{-1.1cm}\psi_-\ddot{\psi}_-+\dot{\psi}_-^{2}+\dot{\theta}^{2}-4(u_{-,\theta}+w_{-,\theta})\psi_-\dot{\psi}_-\dot{\theta}
\approx0,\n{D10a}\\
&&\hspace{-1.1cm}\psi_-\ddot{\theta}-2(u_{-,\theta}+w_{-,\theta})\psi_-(\dot{\psi}_-^{2}+\dot{\theta}^{2})+2\dot{\psi}_-\dot{\theta}\approx0,\n{D10b}\\
&&\hspace{-1.1cm}e^{-4(u_-(\theta)+w_-(\theta)-u_1-w_1)}\psi_-^{2}(\dot{\psi}_-^{2}
-\dot{\theta}^{2})\approx16\,.\n{D10c}
\ea
According to expression \eq{D8}, the order of approximation in the geodesic equations \eq{D10a}--\eq{D10c}
corresponds to the order of approximation of the metric \eq{6.1}.

We use the shift freedom of the proper time $\tau$ to set $\tau=0$ at the singularity for each of the ``radial'' geodesics approaching the singularity (see footnote 4). The point $\tau=0$ is a singular point of equations \eq{D10a}-\eq{D10c}. To find an approximate solution to the geodesic equations near the singular point, one can apply the method of asymptotic splittings described in \cite{CotBar}. A ``radial'' geodesic approaching the singularity is uniquely determined by the limiting value $\theta=\theta_{0}$ at $\tau=0$. The asymptotic expansions of $\psi_-$ and $\theta$ near $\tau=0$ have the following form:
\ba
&&\hspace{-0.5cm}\psi_-=\,2\sqrt{2}\,\tilde{\tau}^{1/2}+\frac{3}{\sqrt{2}}f_{,\theta}^2(\theta_{0})\,\tilde{\tau}^{3/2}+{\cal O}(\tilde{\tau}^{5/2})\,,\n{D11a}\\
&&\hspace{-0.2cm}\theta=\,\theta_{0}+ 2f_{,\theta}(\theta_{0})\,\tilde{\tau}+{\cal O}(\tilde{\tau}^{2})\,,\n{D11b}
\ea
where $\tilde{\tau}=e^{f(\theta_{0})}\tau$ and $f(\theta)=2(u_-(\theta)+w_-(\theta)-u_1-w_1)$.

\end{document}